# In-situ surface porosity prediction in hybrid-directed energy deposition process using explainable multimodal sensor fusion

Adithyaa Karthikeyan, Himanshu Balhara, Abhishek Hanchate, Andreas K. Lianos, Satish T.S. Bukkapatnam

*Wm Michael Barnes'64 Department of Industrial and Systems Engineering, Texas A&M University, College Station, TX 77843, USA*

## ABSTRACT

In this paper, we relate time–frequency patterns of acoustic emission (AE) and other multimodal sensor data with pore formations in a hybrid directed energy deposition (DED) printing process. The multimodal data streams are aligned and synchronized at 0.1 msec temporal resolution and aligned to 0.5 mm spatial resolution. An explainable artificial intelligence (XAI) analysis using local interpretable model-agnostic explanations (LIME) indicates that specific high-frequency waveform signatures of AE can be attributed to two major pathways of pore and void formation in a DED process, namely, spatter events and insufficient overlap between adjacent printing tracks from low heat input. This approach opens an exciting possibility for real-time detection of pores in localized sub-mm surface elements – *surfels*, during the DED printing process. The experimental study collected synchronized multimodal data from multiple sensors, including force, AE, vibration, and temperature, during printing and machining of 316L material samples on a hybrid DED printer (Texas A&M University's Optomec–LENS® MTS 500). A deep convolution neural network (CNN) classifier was used to identify the possible presence of pores on every surfel, based on the time–frequency patterns of the multimodal data collected during the process chain. The results indicate that the signals collected during DED printing are more sensitive for detecting porosity in surfels than signals from machining. Consistent explanations drawn from subsequent LIME analysis indicate that the energy captured in high-frequency AE waveforms is ~37% lower for surfels containing pores than for nonporous surfels. This reduction in energy is in consonance with the relatively lower levels of laser–material interaction in the melt pool and the consequent insufficient fusion and poor overlap between adjacent printing tracks. Porous surfels with prevalent spatter events during printing had ~27% higher energy content in the high-frequency AE band than surfels with lack of fusion voids. These results further open avenues for investigating the time–frequency signatures of AE and other signals to discern the diverse pathways of pore and other defect formation in a DED process.

# 1. INTRODUCTION

The global metal 3D printing market is projected to reach USD 10.91 billion by 2030, with a substantial compound annual growth rate (CAGR) of 31.0 % [1]. Within the realm of metal additive manufacturing processes, DED is increasingly considered in the industry for printing functionally graded alloys, metal–matrix composites, coatings, and refurbishment [2]. It employs a concentrated energy source, usually a laser, that converges with metal powder streams at a common focal point in the presence of an inert shield gas. Part accuracy and finish (due to the layering effect), residual stresses, and defect rates, especially of porosity are the key barriers to their wider industrial adoption [3]. In particular, pores and void formations have a detrimental impact on structural integrity, part density reduction, and stress concentration. Consequently, the prevention and mitigation of these anomalies are imperative for achieving high-quality, defect-free parts in additive manufacturing processes.

Pores and voids in DED processes exhibit stochastic formation patterns spread across a size range of $10^{-5}$-$10^{-3}$ m [4-6]. These defects originate from multiple physical mechanisms, and occur at diverse locations, including the surface, subsurface, or between deposited layers [7, 8]. Several factors, including process parameters, powder particle characteristics, and variables affecting thermomechanical coupling, influence the pore formation [9]. Low laser power often leads to incomplete melting of the feedstock (metal powder), resulting in pores, shrinkage voids, and other lack of fusion defects [10]. High scan speeds can result in irregular pores, along with balling, where the liquid scan track may break down and produce smaller spherical particles. High laser power with slow scan speed can cause the vaporized metal to exert a recoil pressure that pushes down the melt pool surface, forming keyhole pores [11]. Keyhole pores are predominantly observed in laser powder bed fusion (LBPF) processes and are less common in DED due to the latter's utilization of a larger laser spot size and lower energy density [12]. The prevalence of Marangoni flow in the melt pool can also create a depression of the liquid surface beneath the laser beam, leading to pores [7]. Furthermore, the utilization of argon gas-atomized powders in DED introduces a significant source of pore formation. Argon microbubbles often encapsulate powder particles during gas atomization and are released into the molten pool during melting. As these bubbles coalesce and grow beyond a critical diameter of ~120 μm, the resultant buoyancy forces exceed the opposing Marangoni shear forces, propelling the bubbles to the surface of the melt pool where they burst, causing potential spattering [12]. Spatters in the DED process can cause stochastic lack of fusion defects and alter the thermomechanics, leading to surface pores [4].

In this study, we relate to two major pathways of pore formation: namely, insufficient overlap between adjacent printing tracks from low heat input, and spatter events. Pertinently, pores and voids resulting from lack of fusion, as shown in Fig. 1(a), are predominantly observed between successive printing scan tracks due to insufficient overlap. These defects are typically oblong with relatively smooth boundaries and have spherical equivalent diameter (SED) greater than 150 μm [13], depending on the process settings and hatch spacing. In contrast, pores originating from factors such as moisture evaporation and spatter are typically more circular, with highly irregular/fractal boundaries. Pores formation size associated with spatter events in the DED depend on the powder particle size and the laser-material interaction (Fig. 1(b)), and preferentially occur along the middle band of the printing tracks [14, 15]. While gas entrapped bubble bursts contribute to spatter formation and can disrupt the stability of melt pool, it is essential to recognize that spatters may also result from alternative sources, such as the deflection of partially melted powder particles from the melt pool's surface [8, 12, 16]

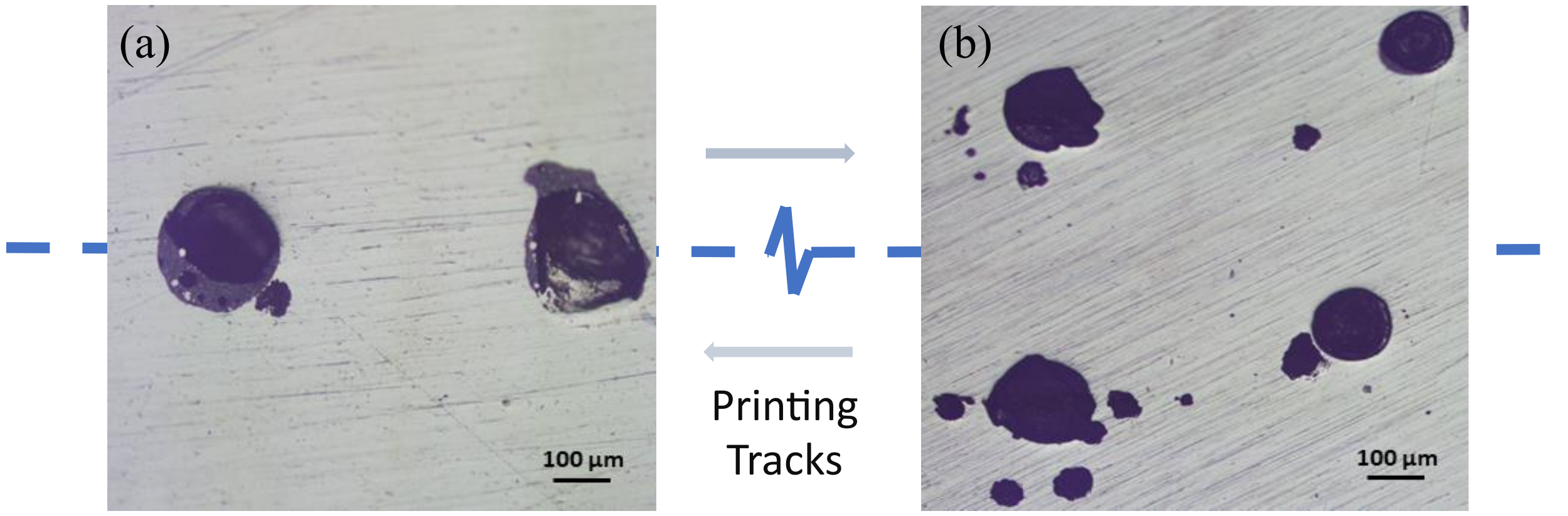

**Fig. 1**. Micrographs of porous surfels captured post milling showing (a) void formations likely due to insufficient overlap between adjacent printing tracks, and (b) pore formations likely due to spatter events.

The diverse and stochastic pore formations have deleterious effects on component quality and performance. Consequently, there is a growing need for real-time, in-situ quality monitoring of the DED process. Sensor-based monitoring enables detection of pores, inference of their underlying formation mechanisms, and the implementation of on-the-fly adjustments to enhance part quality. Prior in-situ monitoring approaches are predominantly vision-based, employing visible light, infrared, X-ray, or multispectral thermal imaging techniques, and a small subset of approaches use sensors such as accelerometers and AE [17-19]. For example, high-speed X-ray imaging is used in a simplified DED-like setup to study pore formation dynamics with micrometer spatial resolution and microsecond temporal resolution [20]. High-speed X-ray imaging and X-ray diffraction techniques allow monitoring of spatter, melt pool dynamics, solidification rates, phase transformation, and pore formation mechanisms due to keyholing at high spatial and temporal resolutions [21]. The industrial scale adoption of in-situ X-ray imaging for DED and other AM processes is constrained by several factors. High costs serve as a primary deterrent, while the need for an elaborate setup requiring frequent adjustments complicates the operations. Further limitations include compromised time-resolution and limited depth of penetration to capture the subsurface and interior pores. Environmental interferences, such as ionizing radiation, material-specific emissivity, and reflectivity issues, further limit its applicability. Conventional ultrasonic and thermography techniques offer cost and time efficiencies [22], but are not well suited to detect pores in real-time, i.e., when a track of the material is being printed.

More recently, several machine learning methods have been employed for real-time pore detection from thermal and optical images of meltpools [11, 23-30] (see Appendix 1). These imaging systems employ delicate and expensive optical arrangements, along with complex data storage and ingestion pipelines (to deal with over 10 GB of data generated per 5 cm of a build track) [31]. In addition, the signal-to-noise ratio of these thermal images is often not high enough to enable accurate pore and defect detection. Alternatively, the high sensitivity of AE sensors, relatively small footprint, and robust and inexpensive hardware offer attractive solutions for real-time in-situ quality monitoring of AM processes. They provide higher temporal resolutions (MHz sampling rates) and faster processing times (1D time-series versus high-dimensional tomography data). AE sensors capture transient elastic stress waves emanating from laser–material interactions and have proven valuable in gathering signatures of complex physical phenomena, including melting, solidification, crack propagation, and pore growth [32-40] (see Appendix 1). Tempelman et al. [38] achieved 93% keyhole porosity prediction accuracies within laser powder bed fusion (LPBF) processes using acoustic emission (AE) signals through first-order statistics, intrinsic mode functions and spectral power of various frequency bands, tested across multiple time windows. In DED however, simultaneous powder material flow rate and its fusion at the laser source's focal point demand high-resolution time-frequency analysis of acoustic signals to accurately discern various process events and ensure uniform deposition and porosity mitigation.

As part of the growing "Industry 4.0" transformation, manufacturers are integrating their physical production and operations with these sensors and smart digital technologies, thereby creating a highly interconnected cyberphysical network to enable efficient and automated production workflows [41]. The increasing affordability and adoption of sensor fusion in manufacturing, where data from multimodal sensors are fused to enhance information veracity, contribute significantly toward understanding the underlying physical phenomena and improving product quality [42-44]. Petrich et al. [45] introduced a supervised multimodal sensor fusion strategy for inter-layer flaw detection in powder bed fusion additive manufacturing, utilizing various sensor modalities, including pre and post laser scan imagery, acoustic and multi-spectral emissions, and insights from scan vector trajectories. Zhang et al. [46] proposed a turning process condition monitoring system utilizing features extracted from power meter and accelerometer to effectively monitor tool wear, chatter and tool chipping. Mohd et al. [47] conducted a survey on multi-modal data fusion for enhancing human-machine interaction in robotic applications by synchronizing natural input methods such as voice, gestures, sensor and haptic feedback, eventually transforming it to a multimedia system output. The authors' previously demonstrated the capability to execute multiple process chains in-situ using Texas A&M's smart hybrid DED machine tool [48]. This smart hybrid machine tool is integrated with diverse small footprint sensors and in-situ imaging capability, allowing synchronous acquisition of 14 channels of multimodal data. This opens up interesting possibilities to track spatiotemporal evolution of the multi-physical states through the entire process chain, and thereby discern porosities at higher spatial resolutions, almost in real-time.

Nevertheless, earlier machine learning methods to predict porosity from various data sources predominantly employ 'black box' models. Although these models may accurately predict the presence and extent of porosity over a limited set of process conditions, they notably lack the explanatory power to discern the underlying physical phenomena. In this context, the adoption of XAI techniques have gained significant momentum [49]. Gawade et al. [50] computed SHAP explanation values for input features including laser speed and laser power, to provide insights into the explanations of layer-wise emission predictions made using a DNN-LSTM framework. Bordekar et al. [51] employed LIME to elucidate the predictions made by a support vector machine (SVM) classifier, offering insights into the pixels containing defects, such as pores or inclusions, in CT scan images of metal additive manufacturing (AM) parts. The authors' recent efforts suggest that eXplainable AI (XAI) approaches, when correctly tuned, can help uncover the physical relationships captured in the powerful machine learning models to help localize pores to high sensitivity and specificity [22, 52, 53].

This is perhaps the first study to employ XAI with multimodal sensor fusion towards on-line real-time detection of pores in localized sub-mm surface elements (henceforth called as *surfels*) of DED-printed 316L parts and to identify plausible pore formation pathways. This study focuses on detecting two manifestations of surface pores: (a) micro-void formations caused by insufficient overlap and fusion between adjacent printing tracks and (b) pore formation due to spatter events during the DED process. In fact, the experimental settings were designed (see Section 2) to induce porosity due to the lack of fusion between adjacent printing tracks (with low heat input) and spattering (with high feed rates); the other pathways of pore formation are less likely. In addition, the designed experiments provide sufficient data to ensure consistent explanations (i.e., discerning of underlying physical mechanisms). They are not typically adequate to provide consistent predictive power over a range of settings and operating conditions. The study's contributions are summarized as follows:

1. Experimental investigations establish that AE signals during the printing cycle capture most information for porosity prediction compared to data from other sensors collected over the process chain (i.e., printing and subsequent machining stages).

2. Inferences from XAI (LIME analysis) help delineate how time-frequency patterns, especially subtle energy variations in high-frequency AE components during DED printing, are mediated by the lack of fusion and spatter pathways of surface porosity. These inferences were further confirmed using thermal melt-pool camera images.
3. Plausible process physics and hence, the type of surface porosity in every sub-mm surface element of hybrid DED-printed 316L components was discerned using multimodal sensor fusion coupled with XAI. Real-time multimodal sensor data captured during the process chain aligned and synchronized to within 0.1 msec accuracy and aligned with 0.5 mm spatial resolution, helped achieve porosity predictions with high accuracy in surfels (69 correct porosity predictions out of 72 surfels).

The remainder of this paper is organized as follows: Section 2 describes the smart hybrid manufacturing implementation at Texas A&M University and outlines the formulation of surface porosity prediction as a binary classification problem. Section 3 explains the mathematical representation of the XAI approach using LIME. Section 4 discusses the performance of various CNN models and provides insights into XAI-fused thermal image data, highlighting the high-frequency waveform signatures of AE pertaining to the two major pathways of pore and void formation. Conclusions are presented in Section 5, which summarizes the key findings and implications of this research.

## 2. EXPERIMENTATION

### 2.1. *Smart Hybrid Manufacturing Implementation*

The Texas A&M smart manufacturing implementation is based around an Optomec LENS® MTS 500 hybrid machine capable of both additive and in-situ machining/finishing operations. It employs a laser head to perform a blown-powder Directed Energy Deposition (DED) process and contains a vertical rotating spindle with an 8-armed turret to perform machining, grinding, and polishing processes. The machine offers a build volume of $350 \times 325 \times 500$ mm$^3$ for additive mode and $500 \times 325 \times 500$ mm$^3$ during machining mode. The hybrid machine is instrumented with various sensors and imaging systems and can simultaneously acquire 14 different data streams [54]. This capability enables real-time monitoring across various processes. Fig. 2(a) and 2(b) illustrate the hybrid machine along with some of the instrumented multimodal sensors. In this work, the primary utility is based only on the accelerometer, acoustic emission (AE), and thermal sensor. Fig. 2(b) highlights these sensors in yellow. The Kistler 500g K-Shear® accelerometer is employed to capture vibrations generated, especially predominant during the machining operations. The Mistras Micro80 miniature AE sensor is used to monitor the transient elastic stress waves released through the process chain. Both sensors are mounted near the build plate attachment under the table of the machine. For the temperature readings, a Stratonics ThermaViz melt pool temperature sensor is utilized to capture and stream the thermal profile during the DED-printing process. Utilizing the capabilities of various multimodal sensors equipped within the DED hybrid machine facilitates deeper understanding of the various physical mechanisms across the manufacturing process chain. Table 1 highlights the capabilities of the various sensors utilized. Fig. 3 illustrates the hybrid-DED setup along with the placement of these sensors.

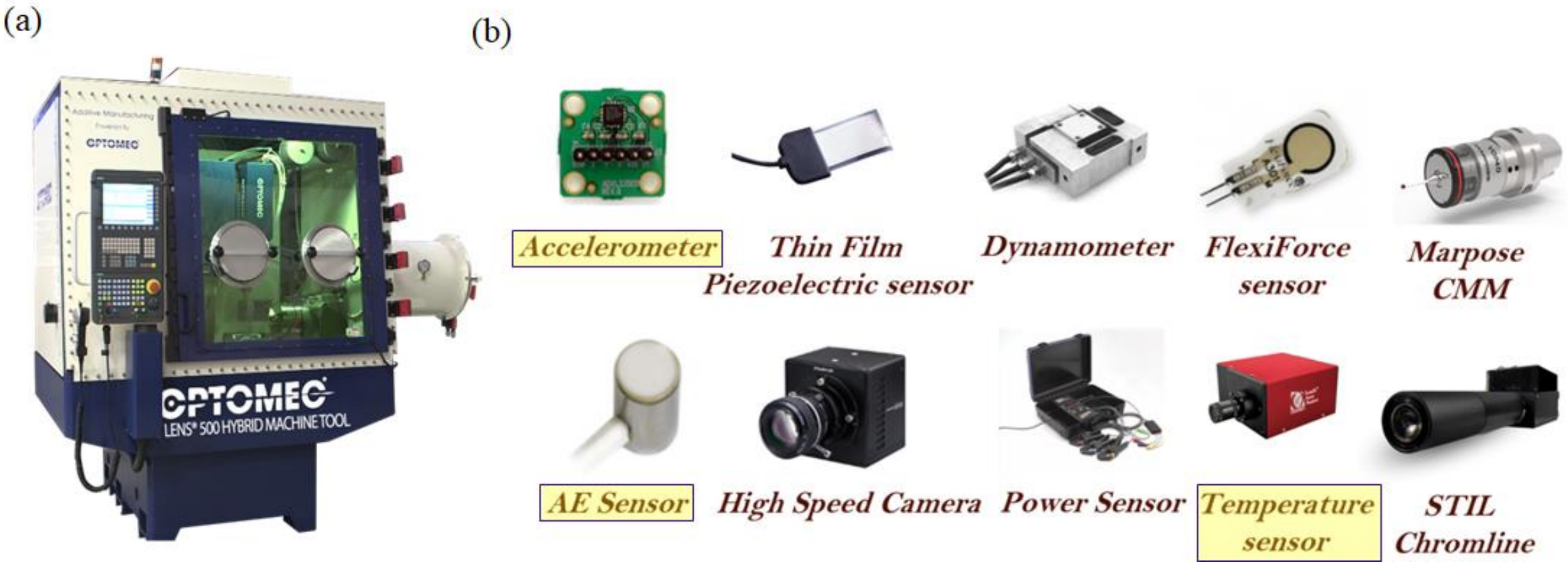

**Fig. 2**. (a) Optomec LENS® MTS 500 DED hybrid machine equipped with (b) an array of multimodal sensors. This study utilizes only accelerometer, acoustic emission and thermal sensors for predicting and explaining the occurrence of porosity in localized surface elements.

| Sensors | Sampling rate | Spatial resolution |
|---|---|---|
| Accelerometer | 10000 Hz | - |
| Acoustic emission | 1 MHz | - |
| Thin film | 1024 Hz | - |
| OPC | 20 Hz | - |
| High speed camera | 6000 FPS | 20 μm |
| Optical camera | 60 FPS | - |
| ThermaViz (Melt pool) | 100 FPS | 14 μm |

**Table 1**. Overview of sensors detailing sampling rate and spatial resolution, with highlighted entries denoting the sensors utilized in this study.

The experiments consisted of printing 316L parts with dimensions $10 \times 10 \times 5$ mm$^3$ on the DED hybrid machine. For this study, AISI Stainless Steel 316L powder particles with diameters in 44–106 μm range were used. To ensure efficient delivery of the powder to the melt pool, the powder gas flow rate was fixed at 6 liter/min. Additionally, a center purge gas flow rate of 25 liter/min was maintained to mitigate oxidation during the printing process. After printing, the top layer was then milled with a depth of cut of 0.072 mm to achieve the desired part geometry and surface finish, with the spindle speed maintained at a constant 3000 rpm. The in-feed during milling was 1.02 mm and the feed rate was 4 mm/s. The hybrid cycle process parameters are listed in Table 2. Fig. 4 shows images of the sample under consideration post printing and milling cycles. It should be noted that milling in-feed is almost twice that of hatch spacing during the printing cycle.

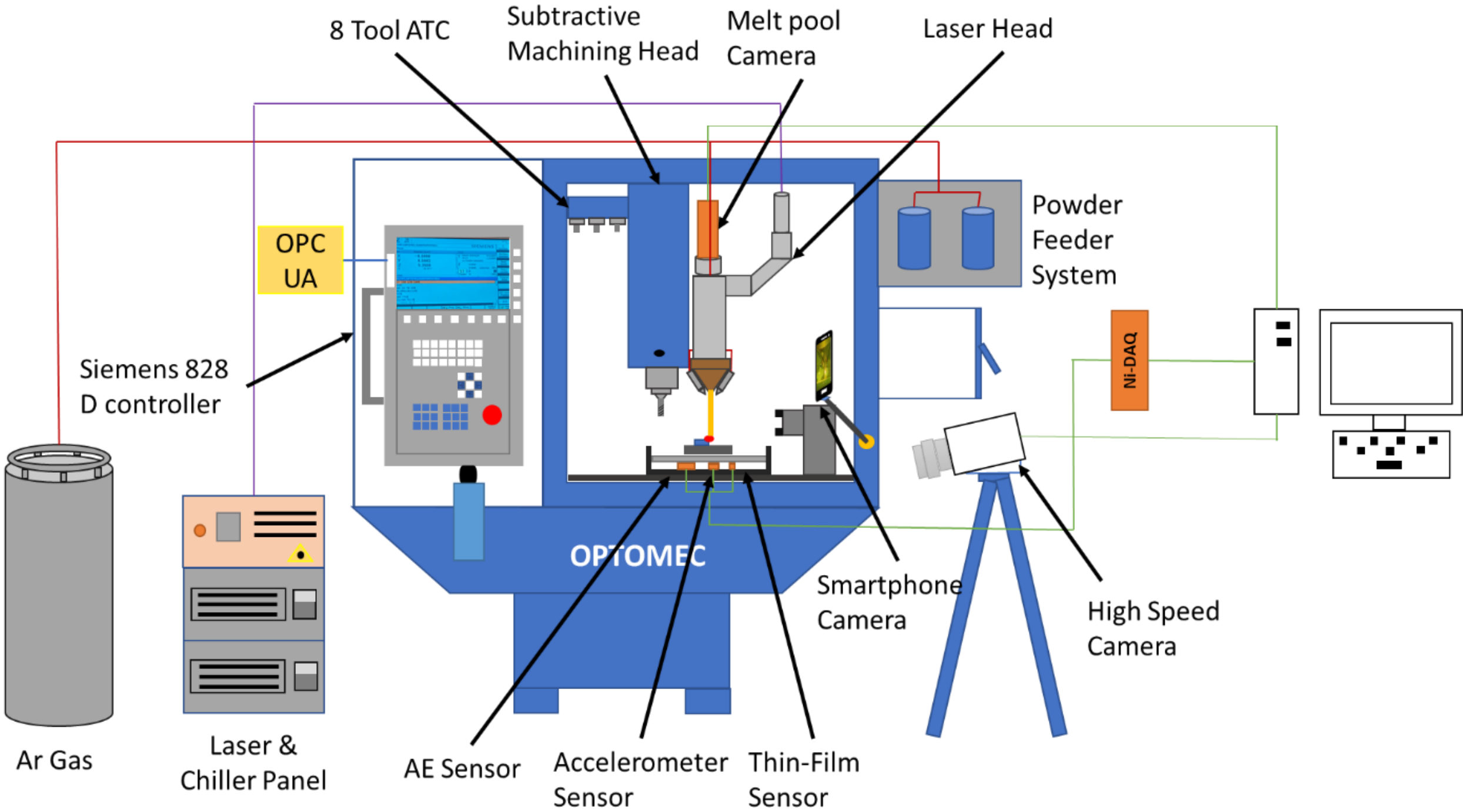

**Fig. 3.** A line schematic of the Hybrid-DED setup, indicating the placement of various sensors. Accelerometer and AE sensors were mechanically coupled to the substrate as they were installed under the tray and thermal sensor was coaxially mounted on top of the laser head.

Excessively large hatch spacing (<50% overlap between tracks) and high scan speed (>15 mm/sec) can further exacerbate the propensity for void formations between adjacent printing tracks under low laser power settings (< 300W) [55]. In this study, the focus is on a sample printed with a low laser power of 250 W and a scan speed of 4 mm/sec.

|  |  |  |
| --- | --- | --- |
| Additive (3D printing) cycle<br>Laser power = 250 W | Hatch spacing | 0.529 mm (0.0208 in) |
| | Scan speed | 4 mm/sec (10 in/min) |
| | Layer thickness | 0.38 mm (0.015 in) |
| | Powder flowrate | 4 rpm |
| End-milling cycle<br>Tool diameter = 9.525 mm (0.375 in) | In-feed | 1.02 mm (0.04 in) |
| | Feed rate | 4 mm/sec (10 in/min) |
| | Depth of cut | 0.076 mm (0.003 in) |
| | Spindle speed | 3000 rpm |

**Table 2**. Hybrid cycle process parameters corresponding to the porous sample under consideration.

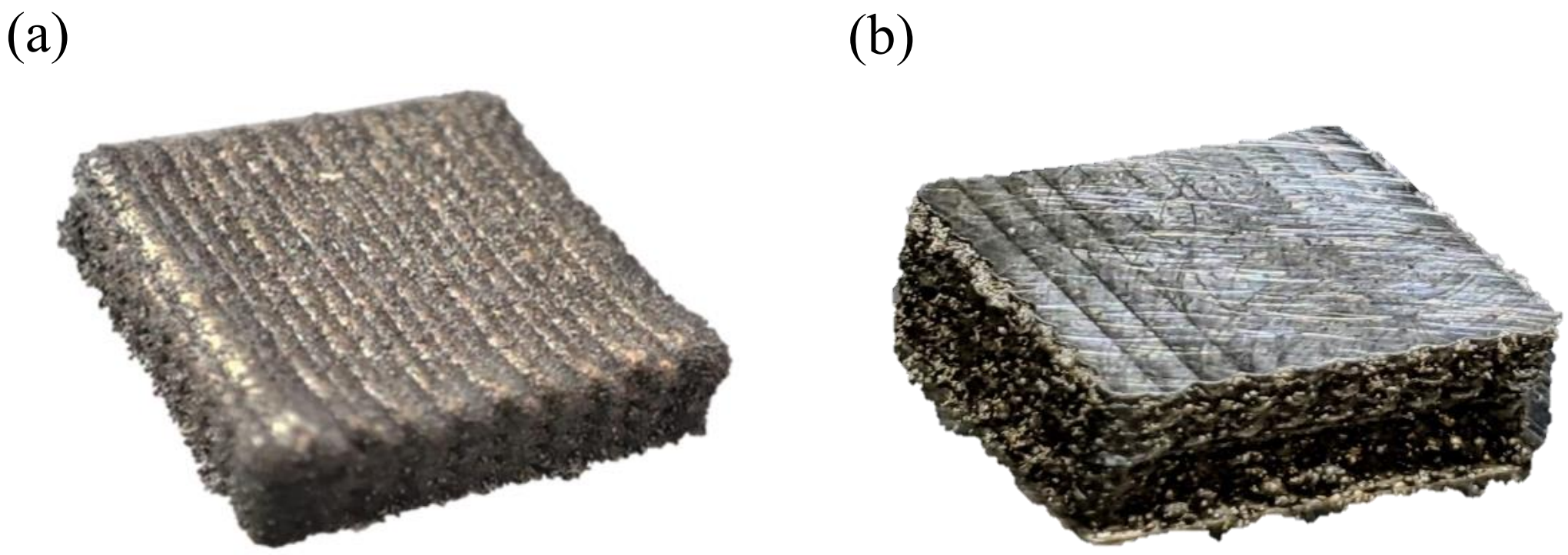

**Fig. 4**. Images of the SS316L sample under study captured using a digital camera (a) post 3D printing and (b) post milling cycle.

## 2.2. *Data Alignment and Synchronization of Multimodal Sensors*

The proposed work is characterized by its multimodal approach to sensing, incorporating data from accelerometers, AE, and temperature (melt pool) sensors. Such multimodal data collection involves diverse data acquisition systems (DAQs), each exhibiting distinct means of digitization with varied sampling rates and resolution. The data fusion aspect in this work is derived upon an innovative process signature-driven multimodal data alignment and synchronization methodology, that allows for high resolution spatiotemporal alignment [56]. The multimodal data streams were aligned and synchronized with a resolution of 0.1 msec for the entire duration of the experiments. Subsequently, the aligned and synchronized data from accelerometer, AE, and temperature sensors were utilized for further analysis and porosity predictions.

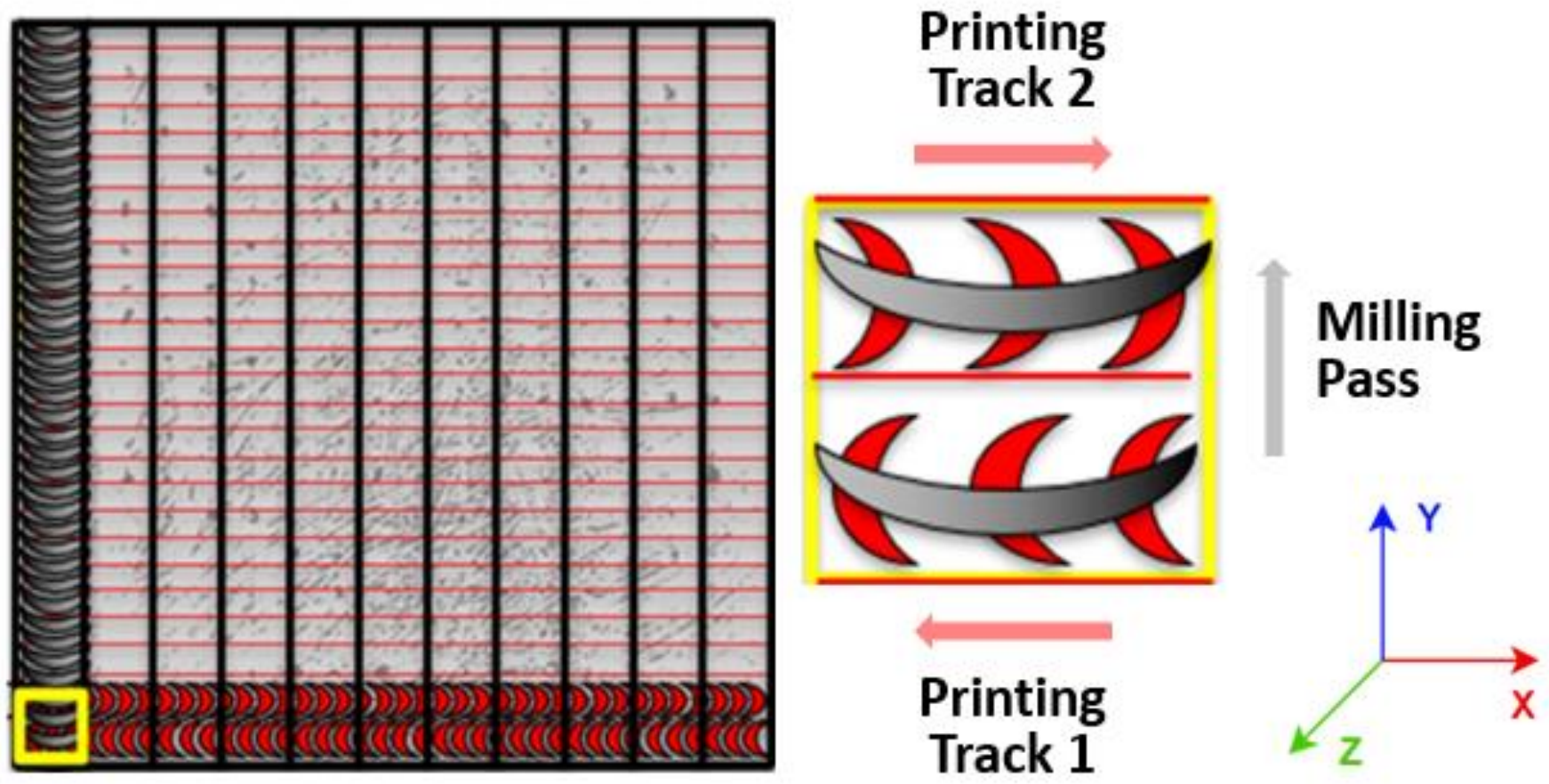

**Fig. 5**. Visualization of the top milled layer featuring an array of surfels containing aligned and synchronized data from multimodal sensors. Each surfel consists of two printing tracks in alternate directions (+X and –X), accompanied by a milling pass orthogonal to the printing direction (+Y axis).

Hanchate *et. al*. [56] also allows for voxelization of multimodal data, wherein the aligned data is segmented and spatially mapped back to segments (voxels) on the physical part itself or its characterization in terms of micrographs. In this work, we employ a similar concept, and the aligned and synchronized data from accelerometer, AE, and temperature sensors is segmented and mapped into surface elements (surfel) of size 1 $mm^2$ on the printed part. Fig. 5 illustrates this in the form of a milled surface of a sample, overlayed with schematics of the printing tracks and milling passes. It also provides a depiction of the segmentation scheme employed to define surfels. Here, each surfel encompasses 2 printing tracks (alternate back-and-forth passes along X-axis) and 1 milling pass along the positive Y-direction.

Data segmentation ensures that each surfel is accurately linked with the corresponding accelerometer and AE aligned signals generated during the printing or milling process of that particular surfel. Specifically, this approach enables the comprehensive capture of both material and information flows at every surfel, effectively tracking their evolution across the entire manufacturing process chain. Here, information based on time–frequency patterns from both accelerometer and AE data channels, represented in the form of their respective spectrograms, collected during the process chain is used to predict the occurrence of porosity in each surfel. The sampling frequency of accelerometer is 10 kHz, while that of AE sensor is 100 kHz. Each accelerometer spectrogram comprises of 129 frequency bands (38.76 Hz bandwidth) and 15 time windows (15 msec width). Similarly, all AE spectrograms consist of 129 frequency bands (387.6 Hz bandwidth) and 15 time windows (15 msec width). The accelerometer and AE signal spectrograms of various stages of the process chain (printing and milling cycles) are stacked together to form a tensor input to deep learning CNN classifiers to identify porous surfels. The high sensitivity and temporal resolution of these sensors, along with their low hardware cost, provide an enticing option to localize the occurrence of porosity, even within surfels.

### 2.3. *Representation of Surfel-wise Porosity using Images*

Porosity in surfels is quantified by measuring the surface area of pores [57], which is estimated from the images of the milled surface using ImageJ software. A binary image labeling scheme is devised wherein images of surfels with cumulative area of pores less than 1% (avg. SED = 36.3 μm) are labelled as non-porous (Fig. 6(a)), and those having an area of pores greater than 2% (avg. SED = 97.2 μm) are labelled as porous surfels (Fig. 6(b)). Fig 6(c) presents a histogram plot representing the percentage area of pores for all surfels, indicating a balanced number of porous and non-porous surfels within the dataset.

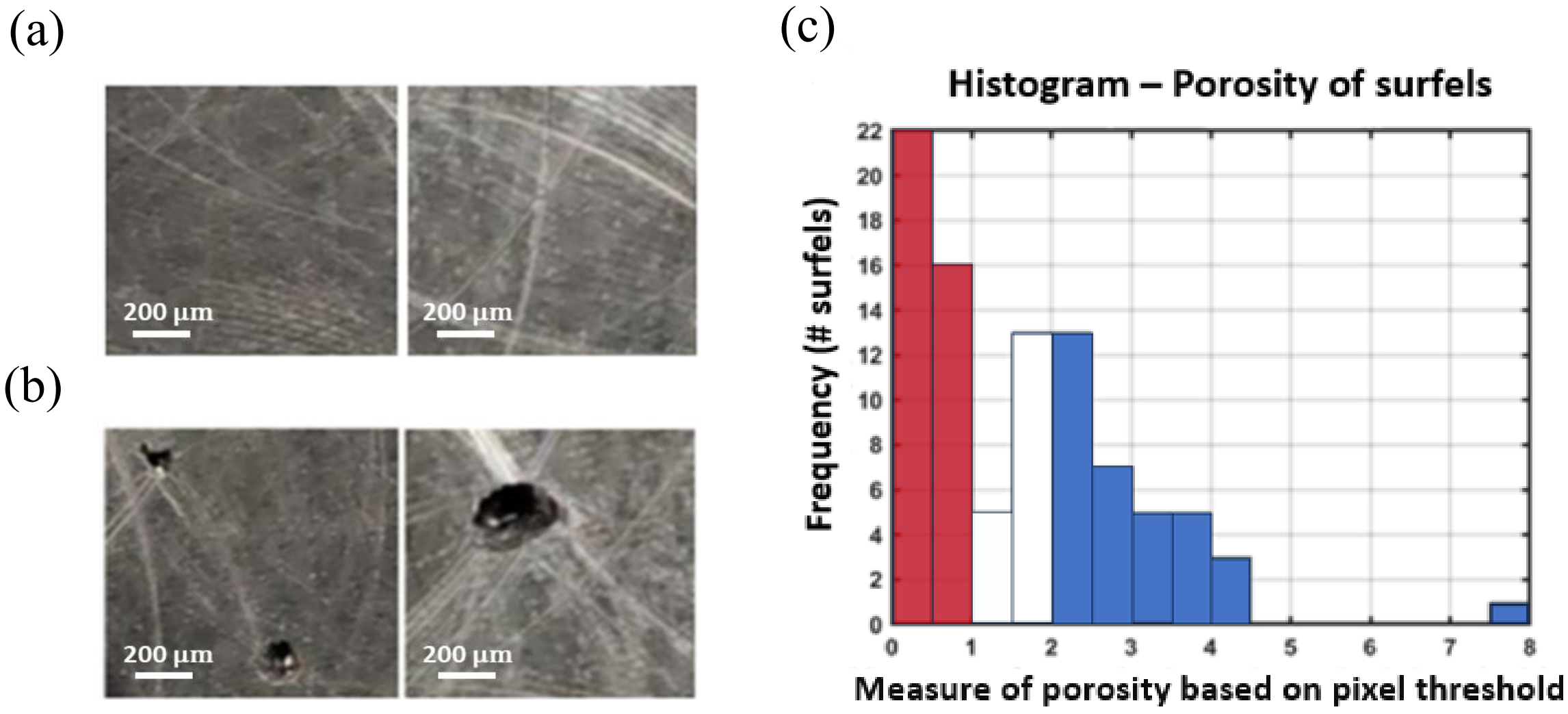

**Fig. 6**. Images of (a) non-porous and (b) porous surfels, in the top layer of the 3D printed part captured post milling cycle (c) histogram of porosity measures (percentage area of pores) in surfels based on pixel thresholding using ImageJ software. Partitioning surfels into non-porous (< 1%) and porous (> 2%) classes helps establish a robust decision boundary for classification.

## 3. XAI AND LOCAL INTERPRETABLE MODEL-AGNOSTIC EXPLANATIONS (LIME)

During the past decade, multiple deep learning models have been applied for real-time process monitoring and anomaly detection in the manufacturing domain, particularly by fusing multimodal data from various sensors. However, the intricate non-linear nature of these deep learning models often lead to challenges in interpreting model predictions and comprehending the impact of various input features. This lack of transparency impedes the ability to gain insight into the relationships encapsulated in these 'black box' models. This limitation reduces the level of trust and confidence that users or stakeholders may have in the predictions made by these models. Therefore, beyond achieving high prediction accuracies using complex deep learning models, it is crucial to explain why the model arrived at a specific prediction outcome for a given instance. This insight, in turn, fosters a deeper understanding of the decision-making process involved within such 'black box' models.

In this context, the adoption of XAI techniques has gained significant momentum. They are designed to augment confidence in 'black box' models by providing insights into the rationale behind specific predictions, (such as determining the presence or absence of a pore in a surfel), based on the sensor data input provided. XAI techniques are generally classified into ante-hoc and post-hoc approaches. Ante-hoc approaches leverage transparent models such as linear regression, decision trees, and modified neural networks, making it easier to retrace the built-in decision process. In practice, they are adopted when the underlying relationships are simple or mostly known, and may not be suitable for capturing high dimensional, complex data models.

Post-hoc approaches such as LIME operate under the assumption that ‘black box’ models exhibit differential behavior across various input space regions. Contrary to providing a global explanation for the complex model across all inputs, LIME focuses on the model’s prediction for a specific instance or a small neighborhood of instances in the input space, providing localized and interpretable explanations. In this context, an explanation involves insights into how different features of the input space affect the model’s prediction. Typically, local surrogates, often linear models, are employed to effectively encapsulate the behavior of ‘black box’ model in proximity to a particular input [52, 58]. Fig. 7 illustrates the LIME concept, using a CNN architecture for classifying defects in 3D printed parts using ultrasound images as a representative example of a ‘black box’ model [22]. The plots to the right in Fig. 6 shows the linear approximations (green) for the predictions on the perturbed dataset (orange) for different inputs. The ‘black box’ model can be locally approximated to the linear model, where the coefficients signify the importance of each feature inferring the model’s prediction. In this study, we extend the implementation of LIME to the case in which the inputs are 3D tensors comprising signal spectrograms from multiple channels. The objective is to identify significant time–frequency bands from each channel that correlate with the occurrence of porosity in each surfel.

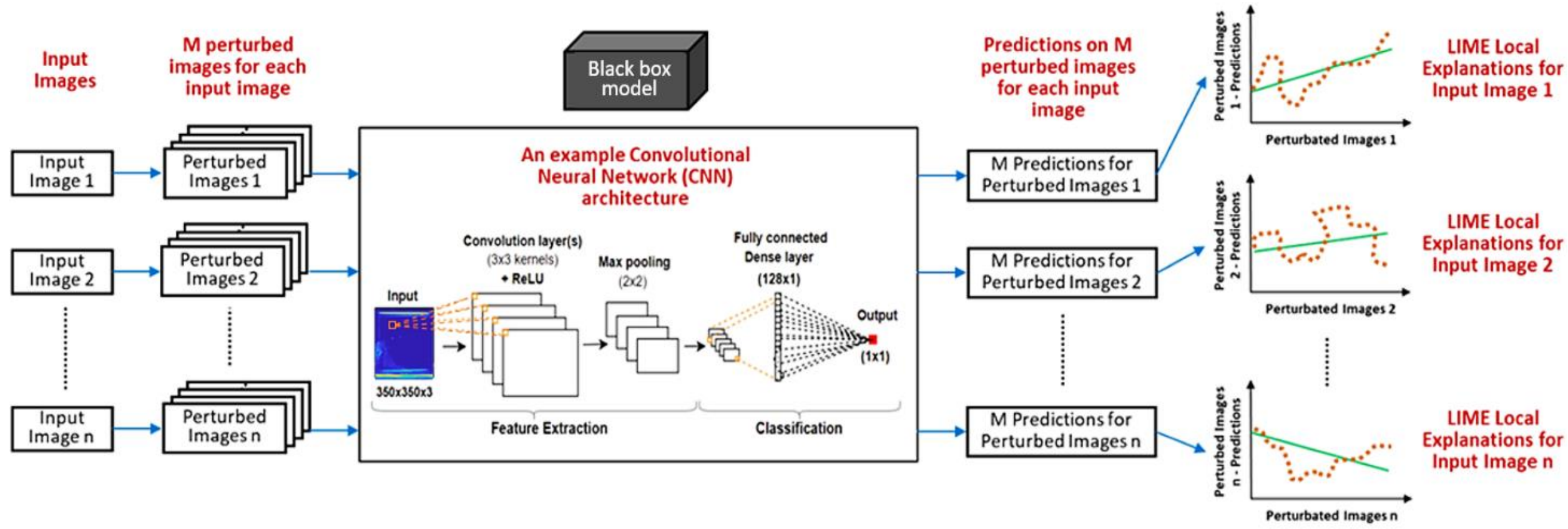

**Fig. 7**. Graphical illustration of the intuition behind LIME for an image-based defect detection.

Mathematically, let the CNN classifier for surfel-wise porosity be denoted as a function $f$, which categorizes the input space comprising signal spectrograms from each surfel into two classes denoting ‘porous’ and ‘non-porous’. To obtain a local explanation of a tensor $T$ in the input space consisting of $N$ channels, LIME first segments each channel $x_i$ (here, a spectrogram) where $(i = 1,2,3, \ldots N)$ into $d$ parts. We define $x$ as the matrix formed by concatenating all channels $x_i$. Unlike an image in which similar pixels can be grouped into segments of many sizes; each component of the spectrogram represents the signal energy over a certain time and frequency. Henceforth, each such component of the spectrogram is referred to as an element. The model under consideration has an input tensor with $N = 4$ signal spectrograms. For each spectrogram, 129 frequency bands capture signal energy across 15 time windows, resulting in a total of $d = 129 \times 15 = 1935$ elements. For every channel $x_i$ in this tensor, a mask $x_i' \in \{1\}^d$ is constructed, i.e., a binary vector with all $d$ elements set to one. All the masks are concatenated to form a single unwrapped mask $x'$ having $N \times d$ elements (see Fig. 8). Subsequently, LIME generates $M$ different perturbation masks $z_j' \in \{0, 1\}^{N \times d}$ where $(j = 1,2,3, \dots M)$ by randomly setting some elements to zero in $x'$ based on a Bernoulli distribution, i.e., Bernoulli(0.5). Each of these masks $z_j'$ is applied to the original matrix $x$ to create perturbed samples $z_j$, $(j = 1,2,3, \dots M)$ in the input space, on which the CNN model $f$ makes predictions. The output layer of the CNN model $f$ uses a sigmoid activation function, resulting in each prediction being a single probability towards the class that is encoded as one. Each perturbed sample $z_j$ is weighted by an index $\pi_x(z_j)$ based on how similar $z_j$ is to $x$. This similarity index is defined using a radial basis kernel as follows:

$$\pi_x(z_j) = exp\left(\frac{-D(x,z_j)^2}{\delta^2}\right) \tag{3.1}$$

where $\delta$ is the kernel width, and $D(x, z_j)$ is calculated as the cosine of the angle between their vector masks $x'$ and $z_j'$. LIME, then trains a linear model $g$, defined by its coefficient vector β*

$$\beta^* = \text{argmin}_\beta \;\; \mathcal{L}(f, \beta, \pi_x) + \lambda\langle\beta, \beta\rangle \tag{3.2}$$

wherein $\mathcal{L}(f, \beta, \pi_x) = \sum_{j=0}^{M} \pi_x(z_j) \left[ f(z_j) - \langle\beta, z_j'\rangle \right]^2$ is the loss function that evaluates how well the linear model $g$ approximates the CNN model $f$ in the vicinity of the original sample $x$. Here $z_0 = x$, $z_0' = x'$, and $f(z_j)$ is the output of the CNN model. Also, $\lambda$ is the ridge regularization term that penalizes the complexity of the linear model. The significance of each element in $x$ can be inferred from the sign and the magnitude of its corresponding $\beta^*$ estimate, indicating whether its presence is positively or negatively important. It is interesting to note that a total of $2^d - 1$ possible perturbations exist to estimate $\beta^*$ about every perturbed spectrum $x$. Considering all possible permutations becomes computationally infeasible. Therefore, a smaller value of $M$ is selected based on the following theorem [1].

**Theorem 1.** *(Perturbed sample size) [58]: The number of perturbed instances M, required to achieve an uncertainty interval width w of feature importance at a user-specified confidence level α can be estimated as follows:*

$$M = 4\varepsilon^2 \,/\, \bar{\pi} * \left[\frac{w}{\phi^{-1}(\alpha)}\right]^2 \tag{3.3}$$

*where* $\bar{\pi} = \sum_{i=0}^{J} \pi_x(z_i)/J$ *is the average weight estimated from an initial set of J perturbed instances,* $\varepsilon^2$

[1] For reader's convenience, the mathematical equations describing LIME methodology (equations 3.1 to 3.3) and Theorem 1 have been provided verbatim from our previous work [22].

*is the empirical sum of squared errors between LIME model and f, weighted by* $\pi_x(z_i)$ *for* $i = 1, 2 \dots J$ *and* $\phi^{-1}(\alpha)$ *is the two-tailed inverse normal cumulative distribution function at confidence level* $\alpha$.

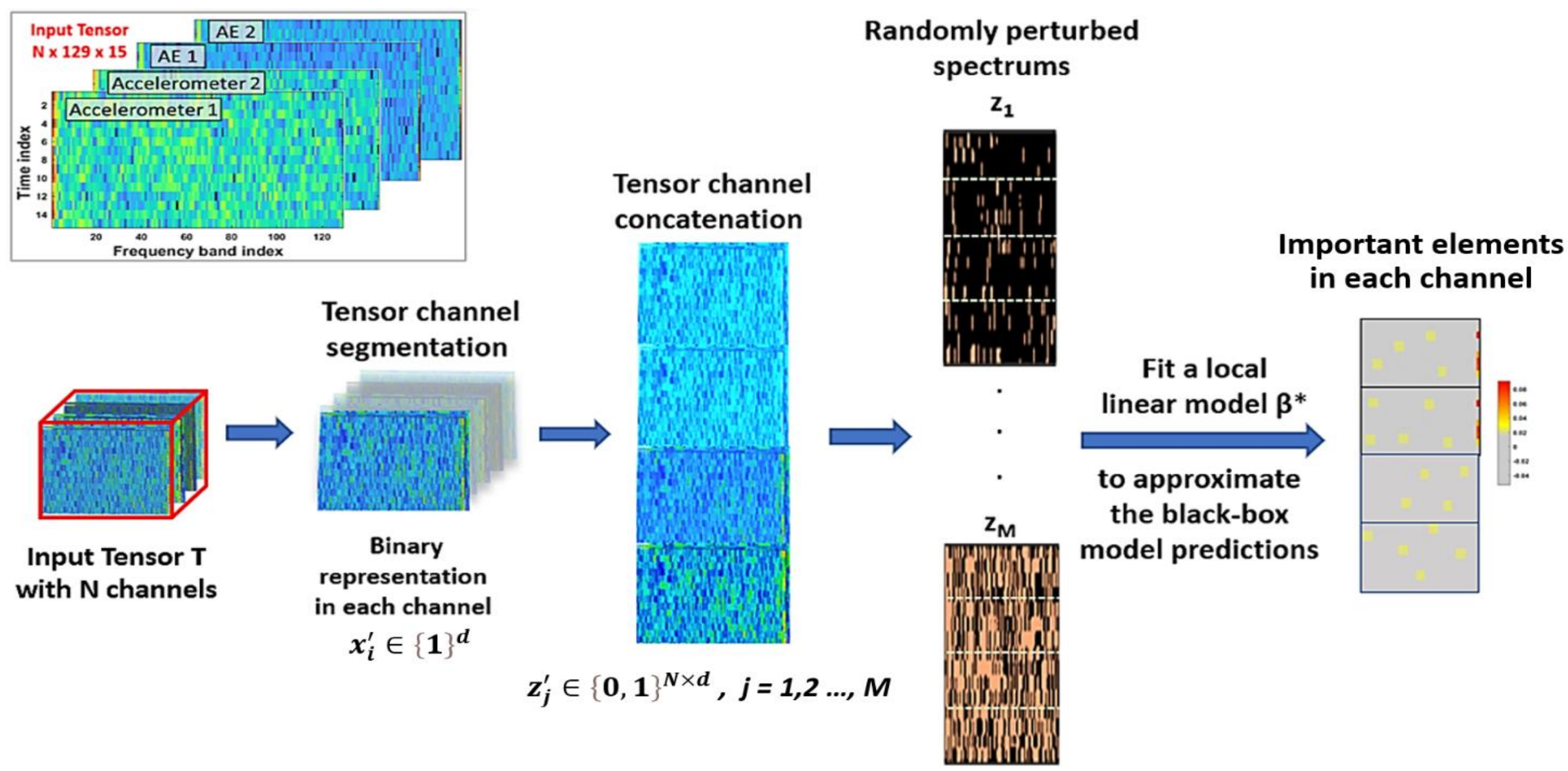

**Fig. 8**. An illustration of the LIME scheme with input features (spectrograms) stacked as a tensor. LIME uses a linear interpretable model to locally approximate the CNN model and helps identify important elements in each spectrogram that contribute towards the CNN model prediction.

Theorem 1 serves as a guiding framework for balancing the trade-off between the number of perturbations required and the uncertainty associated with the explanations. The number of perturbed instances $M$ must be sufficiently large to maintain a small uncertainty width $w$, and at the same time should not be too large to pose a computational challenge. Apart from $M$, the ridge regularization term $\lambda$ should be judiciously chosen [53] — Excessively large $\lambda$ can yield inaccurate local models, and small $\lambda$ can introduce spurious local relationships. Another challenge associated with LIME is the need to consolidate local explanations and derive a global explanation consistent across the input space.

The significance of energy contained in specific frequency bands for certain porous surfels may not hold the same importance for other porous surfels. Such inconsistencies in explanations can be attributed to multiple factors. First, the explanations formulated by the local linear models are derived from a subset of all possible input perturbations generated randomly. Therefore, a different set of perturbations could yield a different explanation altogether. Second, it is possible that linear models may fail to capture the high non-linearity that exists within certain local input neighborhoods. This essentially means that the local linear model may not accurately represent the behavior of 'black box' model in the region of interest, leading to underfitting. Therefore, to achieve consistent global explanations, LIME explanations need to be considered only when the linear model adequately reflects the 'black box' model's behavior locally. The accuracy of linear model $g$ should be high as it attempts to map the perturbed images $z_i$ to $f(z_i) \in (0,1)$. By employing uniform segmentation across different channel frequency bands and considering LIME explanations with $R^2$ scores above 0.75, a measure of global importance can be derived for energy contained in frequency bands from different channels. This measure can elucidate porosity occurrence in surfels and mitigate risk of drawing incorrect explanations from coincidental and spurious relationships that the deep learning models may establish with relatively small training data [53, 59].

## 4. RESULTS AND DISCUSSION

Three different deep learning CNN models were developed to classify each surfel as porous or non-porous, each based on a particular combination of sensor data captured at various stages of the process chain (see Table 3). The input for all the models is a tensor of dimension $N \times 129 \times 15$, where $N$ represents the number of data channels. There are 4 data channels associated with printing cycles (Model 1), 2 channels corresponding to milling cycles, and a total of 6 channels when both printing and milling cycles are considered (Model 3).

| Model | Number of data channels (N) | CNN Model (Input features) |
|---|---|---|
| 1 | 4 | Printing cycles only |
| 2 | 2 | Milling cycle only |
| 3 | 6 | Both printing and milling cycles combined |

**Table 3**. Input data channels used in each CNN model.

For Model 1, the input tensor for each surfel is of dimension $4 \times 129 \times 15$, with the two sensors (i.e., accelerometer and AE) capturing data from both the printing tracks. Model 2 was developed exlusively using accelerometer and AE spectrograms captured during the milling cycle. Here, the input tensor for each surfel is of dimension $2 \times 129 \times 15$. Model 3 utilizes spectrograms from both printing and milling cycles, resulting in an input tensor of dimension $6 \times 129 \times 15$. To minimize bias in data splitting and ensure the robust generalization of each model's performance on the testing dataset, a k-fold cross-validation is performed for each model. The entire dataset was randomly split into training and testing in an 8:1 ratio. An 8-fold cross-validation was conducted on the training set to ensure equitable data samples between the validation and test sets. This translated to 56 surfels being utilized for training and 8 surfels each for validation and testing in each fold. Depending on the value of $N$, the input for the CNN models differs in size but follows a $N \times 129 \times 15$ structure in a tensorized form.

Overall, the study aimed to investigate whether signals obtained during printing cycles exhibit distinct signatures of pores compared to those during milling. The rationale behind predicting porosity in the top layer during the milling cycle is to evaluate the effectiveness of identifying the absence of material within these pores, which can be discerned through sensor data such as vibration or acoustic patterns generated during cutting. Methods used for predicting porosity in 3D printed bulk parts may encounter challenges in precisely attributing a pore to a specific layer, arising from issues during the printing of that particular layer or the effects of melting and solidification in subsequently printed layers. In this context, real-time prediction of pores present in the top layer helps address this concern.

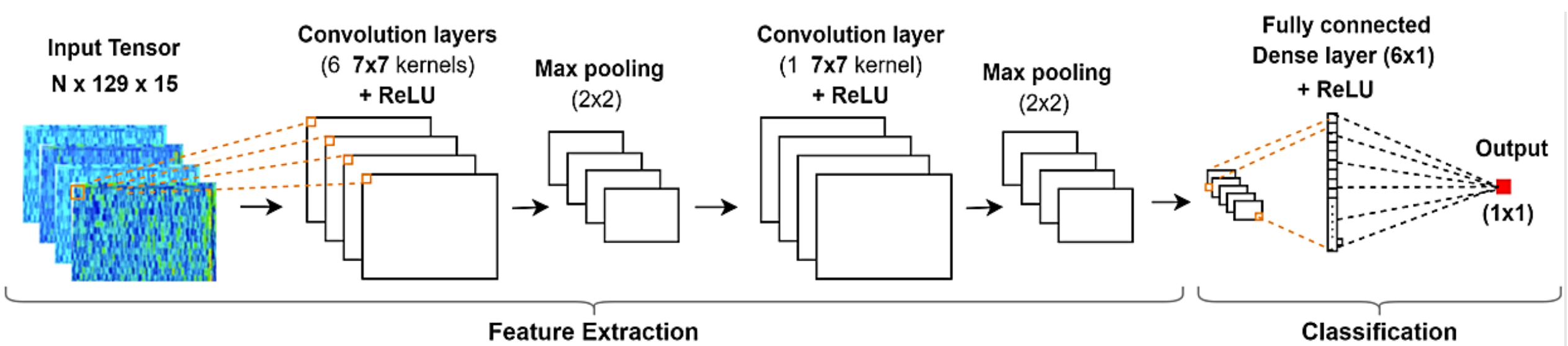

**Fig. 9**. The CNN architecture employed for predicting surfel-wise porosity with accelerometer and AE signal spectrograms integrated into a tensor.

The CNN architecture employed for all models, as summarized in Table 3, consists of two convolution layers (see Fig. 9). The first layer has 6 filters ($7 \times 7$ kernels), and the second layer has 1 filter ($7 \times 7$ kernels). After each convolution layer, the Rectified Linear Unit (ReLU) activation function is applied, accompanied by a max pooling layer of size $2 \times 2$. Subsequently, the network includes a fully connected dense layer consisting of 6 neurons, followed by the output layer with a sigmoid activation function. Upon evaluating various state-of-the-art optimizers, the stochastic gradient-descent (SGD) optimizer was chosen for its computational efficiency in minimizing loss. The SGD optimizer was configured with a learning rate of 0.001.

### 4.1. *Model Prediction Results*

Fig. 10 shows the box plots indicating the training, validation, and testing accuracies from the 8-fold cross-validation of the three models. The training accuracies of the best fold in each of the three models developed were found to be 100%, implying that all 56 surfels were correctly classified as porous or non-porous. The testing accuracy of Model 1 (employing spectrograms from printing cycles) was 87.5% and that of Model 2 (employing spectrograms from milling cycles) was 62.5%. In effect, the information captured by the sensor data collected during printing cycles on a surface is found to be most sensitive for porosity detection. Additionally, the prediction accuracies for Model 1 and Model 3 (employing spectrograms from combined printing and milling cycles) are statistically similar (see Fig.10 (a) and (c)). The observed variations (<5%) in the median validation and testing accuracies in the box plots of Model 1 and Model 3 can be ascribed to the small dataset used in this study.

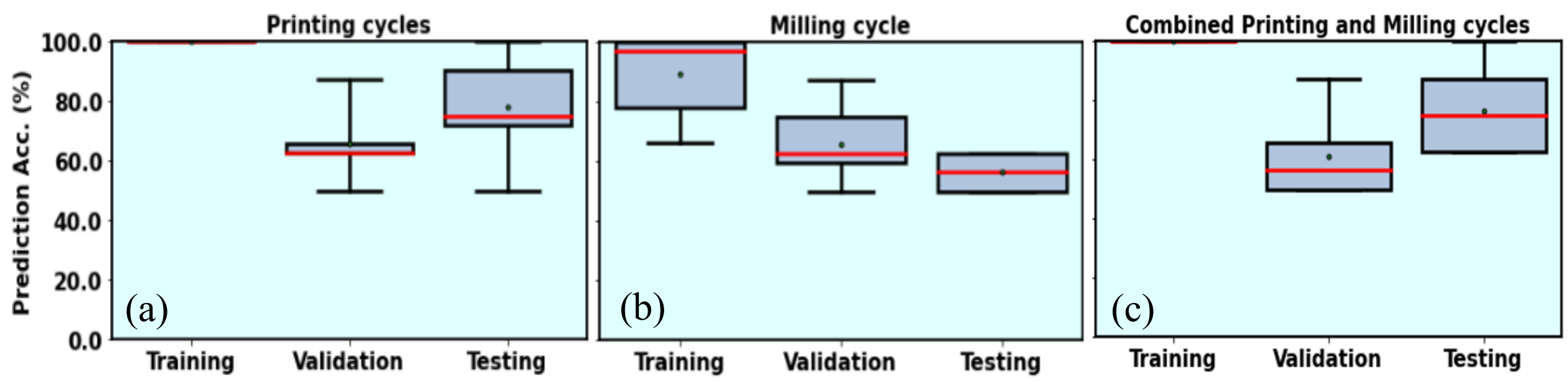

**Fig. 10**. Box plots showing prediction accuracies with training, validation, and testing datasets using signal spectrograms from (a) only printing cycles (Model 1) (b) only milling cycle (Model 2) (c) both printing and milling cycles combined (Model 3).

The confusion matrices presented in Table 4 further delineate the performance of the best classification fold corresponding to Models 1–3, respectively. In the training dataset, no instances of false positives or negatives in identifying a porous surfel were observed. During the validation and testing phases, all porous surfels were correctly classified, with no instances of misclassification. However, the models misclassified three nonporous surfels as porous (false positives). As noted earlier, these models (like the vast majority of the prior porosity prediction models) cannot be deployed for online prediction applications. The classification accuracies (of $\geq$ 87%) indicate that the models can capture the salient physical relationships connecting the signatures of sensor signals with porosity. Section 4.2 presents the inferences drawn from LIME for the model with the highest prediction performance, i.e., best fold from Model 1. Here, the input data is associated with printing cycles and the model demonstrates a Binary Cross Entropy (BCE) measure of less than 0.6 for validation and test datasets, which is the lowest observed across all models under consideration.

| Process Chain | Dataset | Training | | Validation | | Testing | |
|---|---|---|---|---|---|---|---|
| | Prediction class | Porous | Non-porous | Porous | Non-porous | Porous | Non-porous |
| Printing cycle (Model 1) | Porous | 31 | 0 | 3 | 0 | 5 | 0 |
| | Non-porous | 0 | 25 | 2 | 3 | 1 | 2 |
| Milling cycle (Model 2) | Porous | 28 | 0 | 3 | 2 | 3 | 3 |
| | Non-porous | 0 | 28 | 0 | 3 | 0 | 2 |
| Combined cycle (Model 3) | Porous | 27 | 0 | 3 | 3 | 2 | 1 |
| | Non-porous | 0 | 29 | 0 | 2 | 1 | 4 |

**Table 4**. Confusion matrices quantifying the classification performance of the best model-"fold" from (a) printing cycles (b) milling cycle (c) both printing and milling cycles combined.

### 4.2. *XAI-inferred AE features*

For LIME analysis, surfels with high explanation scores ($R^2 > 0.9$) from the first four printing tracks, consisting of 6 correctly classified porous surfels and 6 correctly classified non-porous surfels, are chosen. This choice ensures a sufficiently large sample to assure consistency of explanations while mitigating deriving wrong explanations based on spurious relationships that typically occur in low $R^2$ models. Among the 6 porous surfels considered, posority in 4 are likely due to insufficient overlap between adjacent tracks from low heat input, and the remaining 2 surfels are associated with a high likelihood of pore formation due to persistent spatter events. This labelling of the porous surfels is made based on the shape, size, and location considerations, as discussed in Sections 1 and 2.

Fig. 11 shows the variation of LIME significance values associated with elements from all four channels across the 12 surfels. The figure suggests that the absolute significance $|\beta^*|$of elements from the accelerometer spectrogram was estimated to be less than 0.02 for all the surfels under consideration. As noted in Section 3, the significance $\beta^*$ indicates the influence of that (time-frequency) element in identifying the porosity of the corresponding surfel. Only a few (time-frequency) elements in the 49.61–50 kHz frequency band of AE spectrogram pertaining to both printing tracks (i.e., channels 3 and 4) have significance estimates $|\beta^*| \geq 0.05$. In essence, the high-frequency bands (~ 50kHz) of AE collected during printing bear signatures that capture the pore formations. The ejection of spatters from the melt pool are known to generate elastic waves at high frequency. These distinctive signatures effectively capture variations in acoustic energy released during the processes of solidification and phase transformation, both in the absence and presence of pores. Additionally, pore formations due to spattering are often near-singular events and therefore release energies at high frequencies.

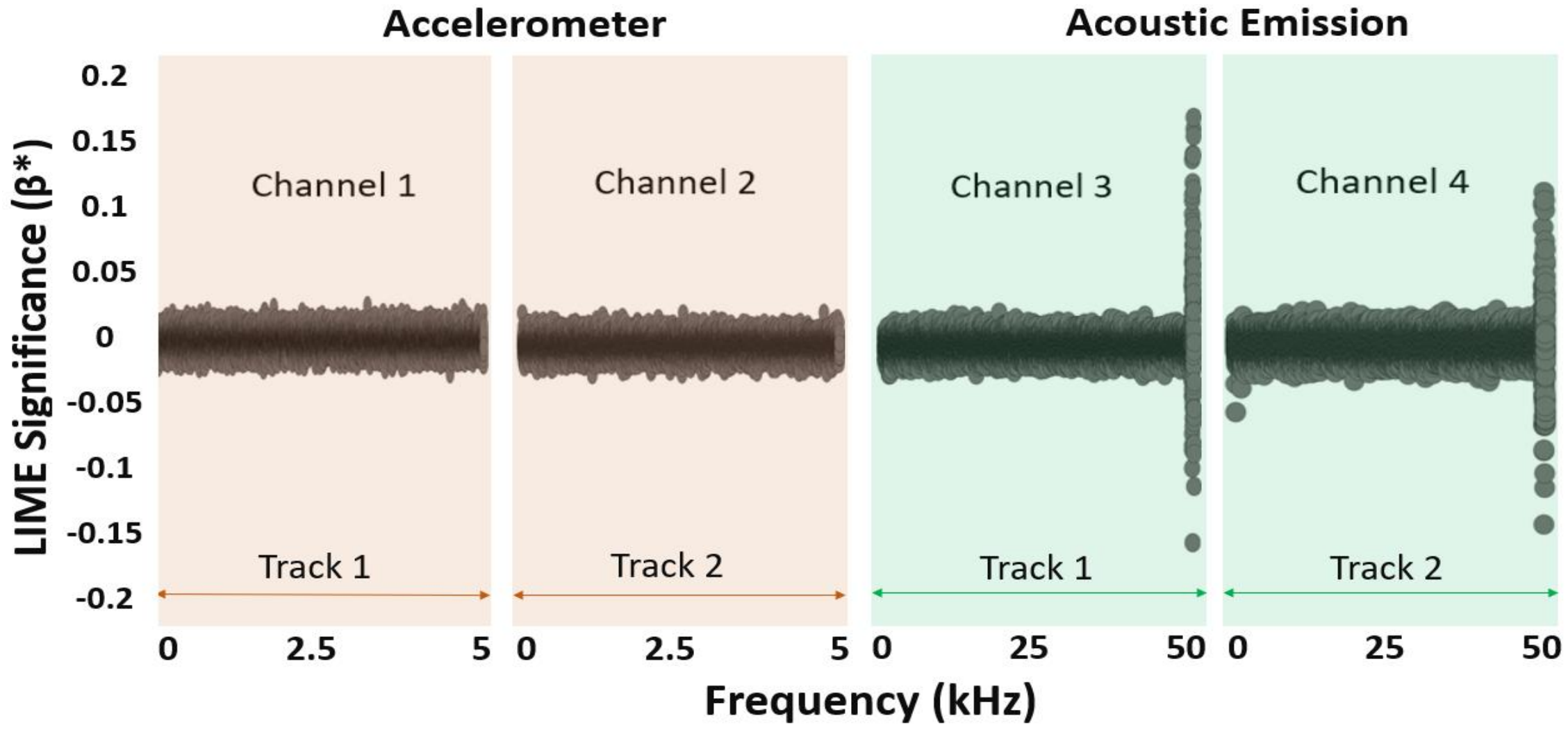

**Fig. 11**. LIME significance of the spectrogram elements contained in 6 representative porous and non-porous surfels under consideration.

Fig. 12. shows the significance of every element in AE spectrograms representing channels 3 and 4 for a correctly classified porous surfel, analyzed using LIME. Here, the color bar quantifies the significance of each element in identifying the porosity of the corresponding surfel. Notably, the high-frequency band 49.61–50 kHz of the AE spectrum (index 129), contains the most significant elements. LIME analysis further reveals that elements within frequency band 11.64 –13.97 kHz (indices 30 – 36) exhibit positive significance (between 0.01 and 0.03) for porous surfels (see Fig. 12). This can be attributed to void formations resulting from insufficient overlap between adjacent tracks. Similar trends supporting the importance of 40kHz band of AE signals have been reported in Pandian *et al*.'s [60] statistical correlation study relating time-frequency characteristics of AE to the lack of fusion, no pores, and keyhole defects in the LPBF process for 316L.

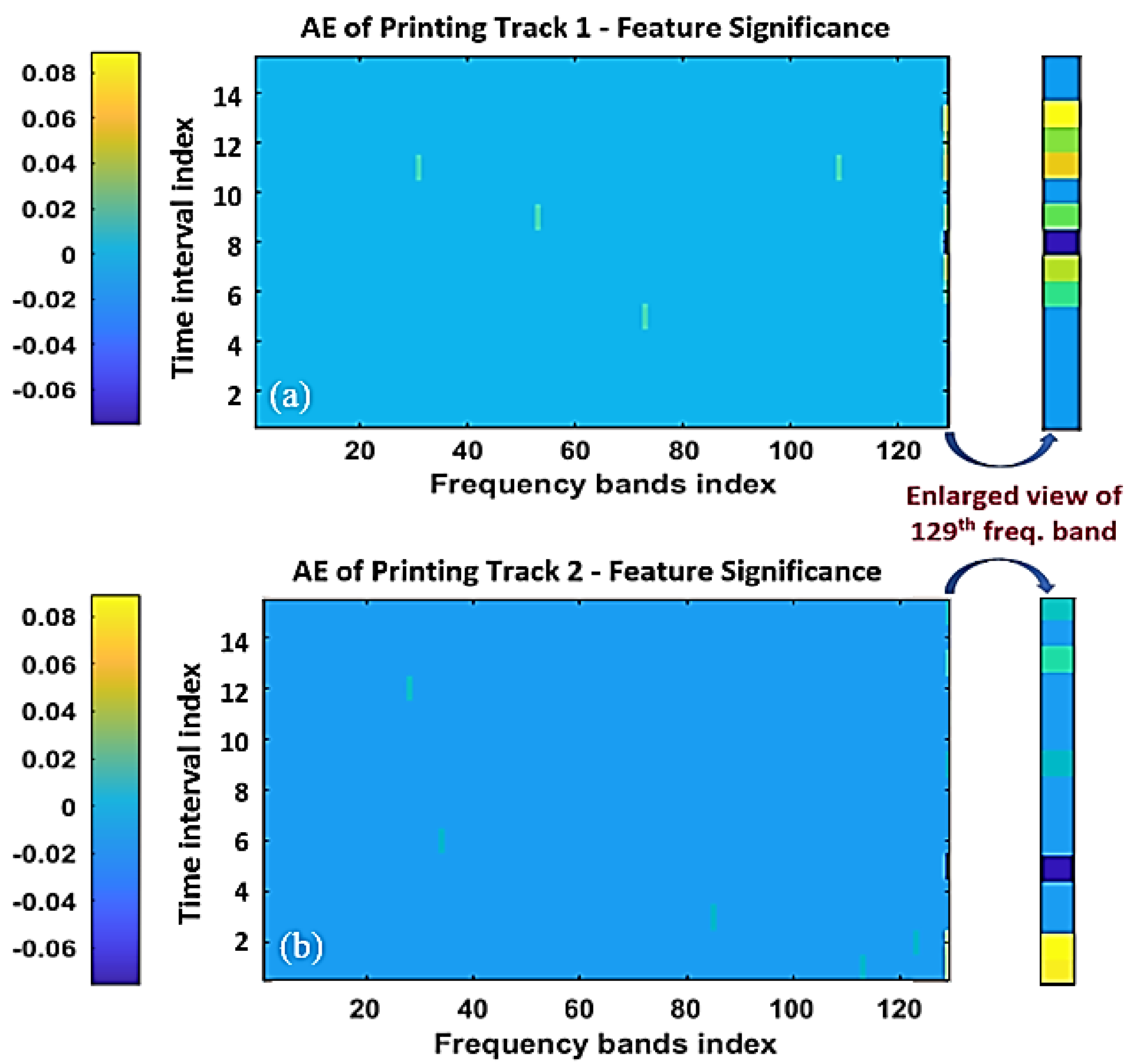

**Fig. 12**. LIME significance of AE spectrogram elements for (a) printing track 1 and (b) printing track 2 in a correctly classified porous surfel.

The box plots in Fig. 13 illustrate the variation in median peak energies contained in high-frequency AE band for the non-porous and the two types of porous surfels. Statistically distinct energy distributions are discernible for the three categories of surfels. Specifically, the median values are observed to be 18.54 dB/Hz for surfels with void formations (attributed to insufficient overlap between adjacent printing tracks from low heat input), 23.61 dB/Hz for surfels with a high likelihood of pore formation due to spatter events, and 28.84 dB/Hz for non-porous surfels. The relatively higher peak energy values observed in the 49.61–50 kHz frequency band of the AE spectrum for non-porous surfels can be attributed to enhanced laser–material interaction within the melt pool. This contrasts with other surfels that exhibit pore formation due to spatter events and voids arising from insufficient overlap between adjacent printing tracks due to low heat input.

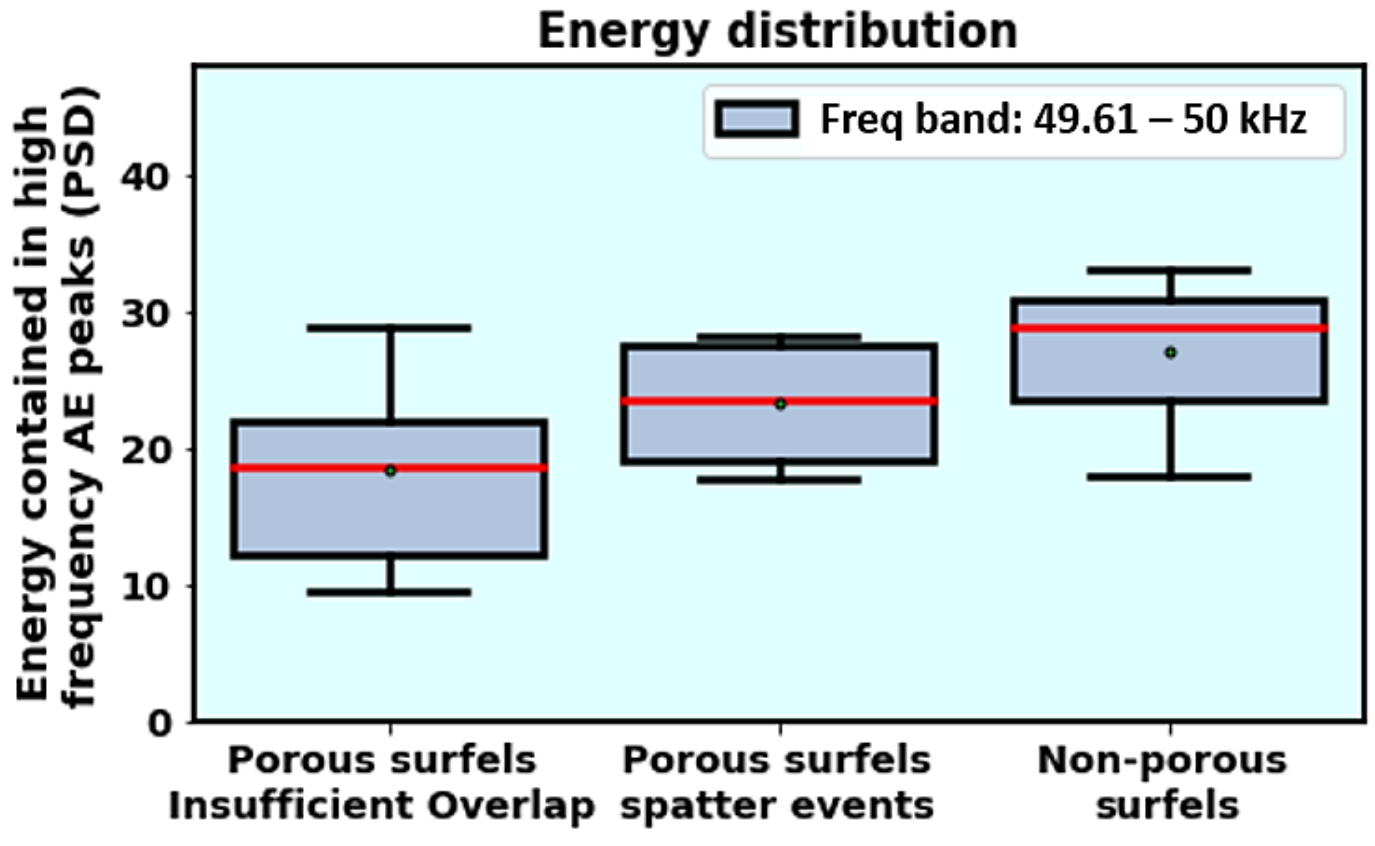

**Fig. 13**. Box plots indicating the variation in median peak energies contained in high-frequency AE band with index 129 for porous surfels attributed to two mechanisms of pore formations, and that of non-porous surfels.

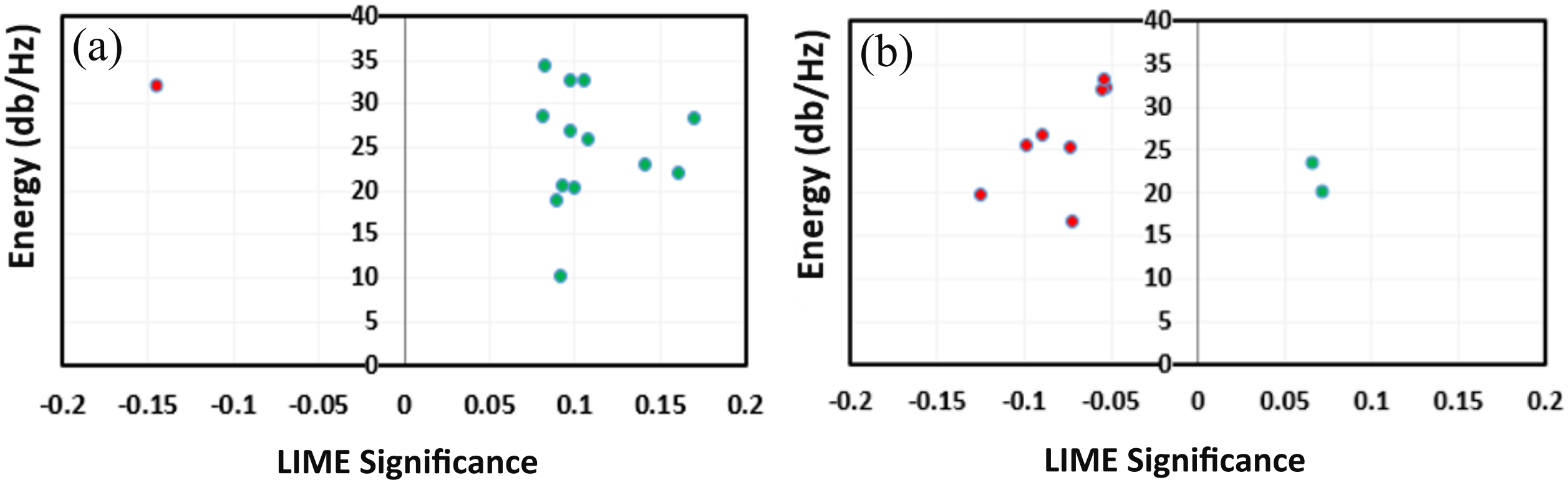

**Fig. 14**. Significant AE Spectrogram elements within the 49.61 – 50 kHz frequency band are depicted, with LIME significance $|\beta^*| \geq 0.05$ for (a) 6 porous and (b) 6 non-porous surfels. Positive significance values contribute towards the presence of pores in the identified porous surfels, while negative significance values contribute towards the absence of pores in the identified non-porous surfels.

Amongst all elements of AE with significance $|\beta^*| \geq 0.05$, (all in the 49.61–50 kHz band), particular attention was directed towards those characterized by elevated levels of energy content. Elements with high energy in this frequency band were found to have positive significance (> 0.08) for porous surfels and negative significance (< – 0.05) for non-porous surfels, as shown in Fig. 14. For the 6 porous surfels, the LIME significance showed positive values within the range of 0.075 and 0.175. Conversely, for the 6 non-porous surfels, the LIME significance consistently remained negative, in (– 0.05, – 0.135) interval. In summary, the energy contained within the high-frequency band (49.61–50 kHz) emerges as the critical signature that governs pore formation during DED process.

### 4.3. *Inferences from Thermal Imaging-AE Sensor Fusion*

Next, we investigated the time-variation of the frequency characteristics, especially to correlate with the short-time (msec) scale spatter events. Fig. 15 illustrates the variation of energy content within high-frequency (49.61–50 kHz band) AE spectrogram elements during a printing cycle spanning four surfels across multiple tracks, of which 2 are porous and 2 are non-porous surfels. The energy contained within the spectrogram elements of the high-frequency AE band exhibits an oscillatory pattern along the printing track. It is noteworthy that non-porous surfels tend to have slightly higher peak energies in this high frequency band in comparision to porous surfels.

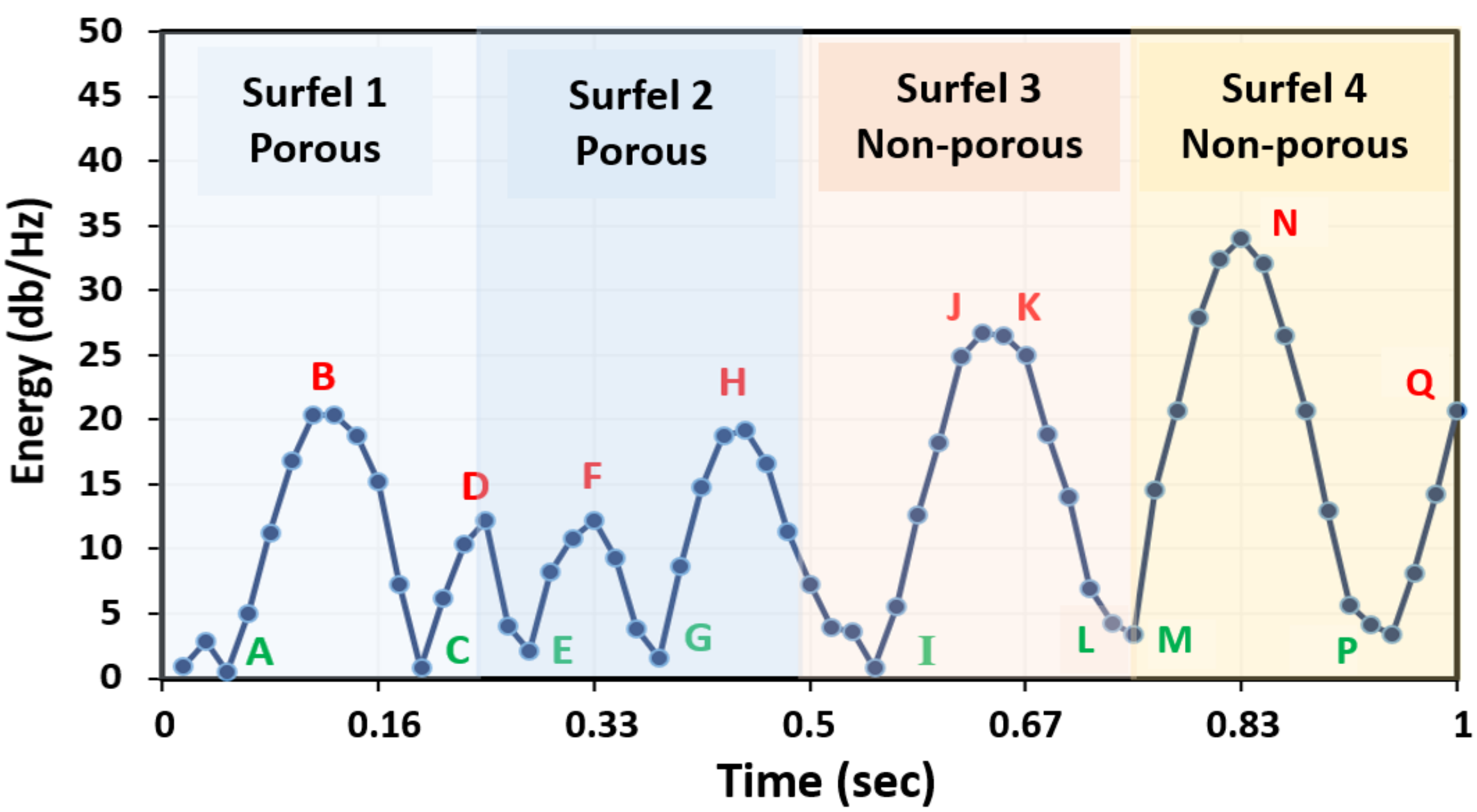

**Fig. 15**. Oscillatory patterns observed in the energy of 49.61 – 50 kHz AE signal spectrogram elements within a printing track containing contiguous porous and non-porous surfels.

Towards further understanding the nature of this oscillatory behavior of high frequency AE elements, we investigated the thermal images taken during the printing of the corresponding surfels. The thermal sensor captures melt pool images at a rate of 30 frames per second. With a scan speed of 4 mm/sec, this results in approximately 7.5 thermal images being captured for every 1 mm of the scanned surface (surfel). Meanwhile, the AE sensor operates at 100,000 Hz. We generate a spectrogram with dimensions of 129 × 15 for each surfel. This translates to 15 time-based elements per surfel across 129 frequency bands in the spectrogram. Consequently, we observe that the 7.5 frames captured by the thermal sensor per surfel correspond to 15 time elements in the AE spectrogram for each surfel. Therefore, each image-frame captured by the melt pool thermal sensor encompasses approximately 2 time elements of the high-frequency AE band. Fig. 16 shows thermograms corresponding to 4 surfels, of which 2 are porous and 2 are non-porous. Remarkably, the thermograms synchronous with the energy peak indices (B, D, F, H, J, K, N, and Q) indicate the presence of spatter events, and those corresponding to valley indices (A, C, E, G, I, L, M and P) indicate the absence of spatter events. This result suggests that the time-variation of the spectral energy for the 49.61-50 kHz AE band is associated with the variations in the spatter activity.

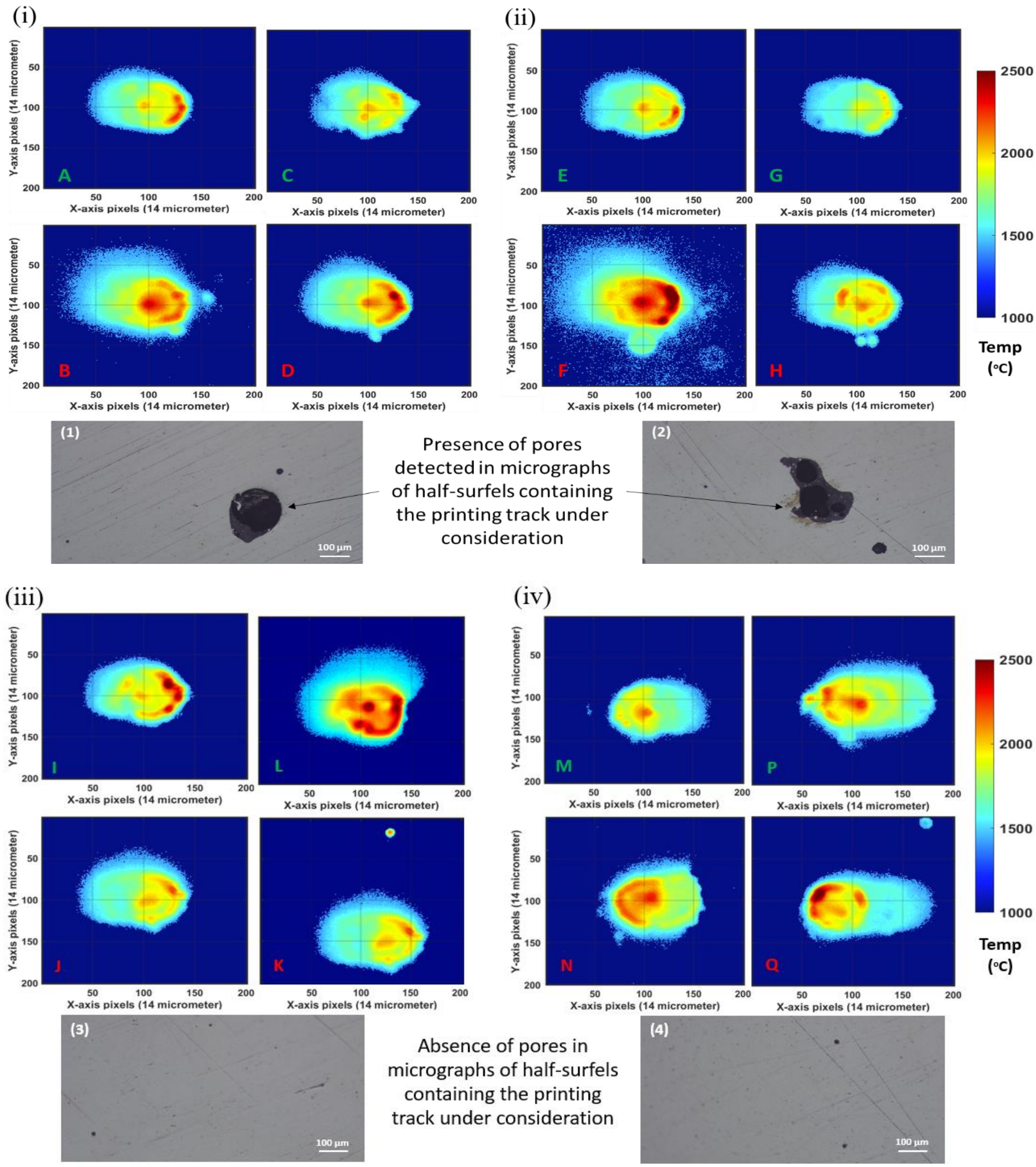

**Fig. 16.** Thermograms and micrographs of surfels 1-4 as indicated in Fig. 15. Besides irregular thermal field patterns, the thermograms of porous surfels show enhanced spatter activity (see image-frames B and F). The micropores visible in surfels corresponding to (iii) and (iv) may have been formed because of trapped gas pockets such as oxygen or nitrogen within the deposited material, as seen by their near circular shape.

To validate the observed correlation between the energy contained in high-frequency AE band and the spatter events captured in melt pool thermograms, 4 complete printing tracks were considered for evaluation. Spatter events were observed during DED printing on most of the surfels. The spatter patterns were captured by high-speed camera as well as the melt pool sensor (ThermaViz® from Stratonics). In fact, among the 46 peaks (crests) observed in the energy values within in high-frequency AE band, 41 of them were identified as containing spatter events. These spatter events were predominantly associated with elevated temperatures within the melt pool. In a couple of instances, false positives occurred, where AE energy was high, but no spatter particles or events were observed in the thermograms. This discrepancy could be attributed to the AE sensor capturing elastic waves from extraneous sources such as the laser beam or table motion system. Similarly, within the 4 printing tracks, 49 instances of valley points (troughs) were observed. Of these instances, 32 were associated with thermograms that exhibited no spatter presence. Additionally, 9 instances fell into the indeterminate category due to the absence of spatter particles and the presence of minor disruptions in the thermograms, primarily stemming from irregular reflections during melt pool capture. These findings have been consolidated in the confusion matrix in Table 5. More pertinently, it is worth noting that the median energy contained in the peaks measures 26.73 dB/Hz, whereas for the valleys, the median energy stands at 2.62 dB/Hz. as depicted in Fig. 17.

| Sensor Fusion (AE / Thermal) | Spatter events | No spatter observed | Indeterminate |
|---|---|---|---|
| Peaks (46) | 41 | 2 | 3 |
| Valleys (49) | 8 | 32 | 9 |

**Table 5**. Confusion matrix highlighting the correlation between energy contained in high-frequency AE spectrogram band and spatter events in melt pool thermograms.

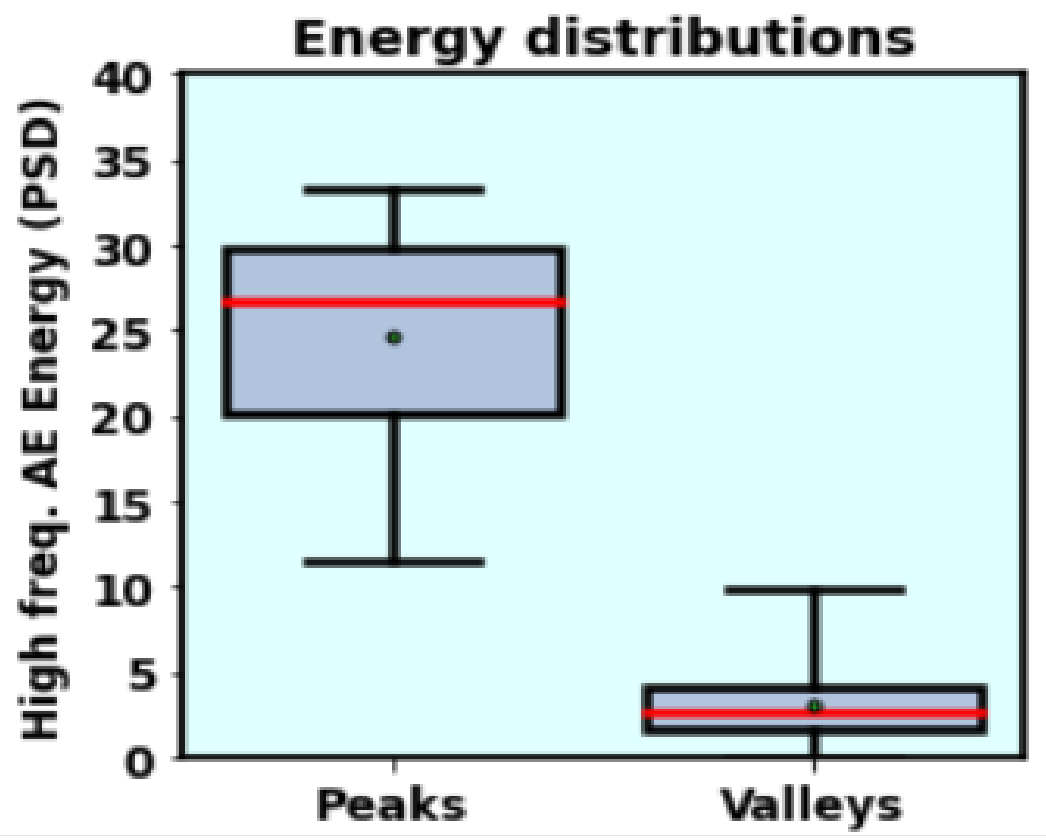

**Fig. 17**. Box plots indicating the differences in energy contained in peaks and valleys of high-frequency AE band (49.61 – 50 kHz) across the 4 printing tracks under consideration.

## 5. CONCLUSIONS AND FUTURE WORK

This experimental study has established key time-frequency signatures of AE that characterize two types of pore formations during a DED process, thereby enabling a high resolution (sub-mm), real-time prediction of porosity within surfels. Salient outcomes from this work are summarized as follows:

1. The high-resolution sensor data synchronization with an accuracy of 0.1 msec and aligned within a high spatial resolution of 0.5 mm, in conjunction with the segmentation of milled surface into surfels (surface elements), provided a precise methodology for tracking both material and information flow at each surfel. This precision was instrumental in enabling surfel-wise porosity predictions and deriving insights into pore formation pathways from AE and thermal melt pool data.
2. A comparative analysis of multiple machine learning models for detecting surfel-wise porosity using sensor data from both printing and milling cycles revealed that the signal data from printing cycles were most effective in identifying the presence of pores in surfels, achieving a prediction test accuracy that exceeded 87%. Moreover, the AE signal from the printing cycles captured most information for porosity prediction compared to other sensors gathered over the process chain (i.e., printing and subsequent machining stages).
3. With the help of XAI, distinctions in peak energy values of the high-frequency (49.61-50kHz) AE band from porous versus non-porous surfels have been identified. These differences correlate with the two observed modes of pore formation, namely, those resulting from spatter events in DED and those due to insufficient overlap between adjacent printing tracks from low heat input. Non-porous surfels were found to have ~37% higher peak energy values in the high-frequency AE band compared to other porous surfels owing to strong laser-material interaction in the melt pool. Similarly, porous surfels showing incessant spatter activity during printing exhibited ~27% higher energy in the 49.61-50kHz AE band compared to surfels with lack of fusion voids.

The robust synchronization of multimodal sensor data, achieved with a high temporal resolution of 0.1 msec and aligned within spatial resolution of 0.5 mm, was instrumental in revealing a fundamental correlation. This alignment allowed us to accurately attribute pores and signals to every surfels, which was essential for correlating them with specific process-related events such as spatter formations, and validating concurrent occurrences through cross-referencing AE data with thermal images. The AE sensor, in particular, has demonstrated its value as a cost-effective and powerful asset for sub-msec scale monitoring and control of manufacturing processes as well as in-situ observation of spattering and other short time-scale phenomena.

Despite the significant findings, this study has certain limitations. First, the primary focus was on elucidating only two specific types of pore formations, and therefore, our analysis may not capture the full spectrum of potential pore variations in the DED process. Expanding the dataset to encompass a wider range of pore types could potentially improve the predictive capabilities of the models. Second, the current models employed in the study are fundamentally based on CNNs. Future research could explore the application of more advanced foundational machine learning models, such as Residual Networks (ResNet), which could potentially yield more accurate predictions.

Nonetheless, this discovery of the AE signatures connoting porosity and spatter events paves the way for future research into time-frequency analysis of AE signals. This exploration can lead to a comprehensive characterization of various spatter and pore formation mechanisms [8], and understanding of laser-material interactions under different process parameters, including laser power, scan speed and powder feed rate. Furthermore, this study opens exciting possibilities for developing mechanistic / computational models that relate DED process phenomena to the various sensor signals, and the possibility of developing monitoring systems based on combining physical understandings with statistical patterns.

## CRediT Authorship Contribution Statement

Adithyaa Karthikeyan: Writing – original draft & review, Methodology, Visualization, Formal analysis, Conceptualization. Himanshu Balhara: Writing – review & editing, Data Curation, Methodology, Formal analysis, Conceptualization. Abhishek Hanchate: Writing – review & editing, Data Curation, Methodology, Visualization. Andreas K. Lianos: Writing – original draft, Data Curation, Methodology, Conceptualization. Satish T.S. Bukkapatnam: Writing – review & editing, Supervision, Project administration, Funding acquisition, Conceptualization.

## Declaration of Competing Interest

The authors declare that they have no known competing financial interests or personal relationships that could have appeared to influence the work reported in this paper.

## Declaration of Generative AI in Scientific Writing

During the preparation of this work the author(s) used ChatGPT developed by Open AI, to improve readability and overall language. After using this tool/service, the author(s) reviewed and edited the content as needed and take(s) full responsibility for the content of the publication. No AI tool was used to analyze and draw insights from data as a part of the research process.

## Data Availability

https://doi.org/10.17632/bfnnn86hhn.1

## Acknowledgements

The authors would like to acknowledge the funding provided by United States National Science Foundation (NSF) under **ECCS-1953694** and **IIS-1849085**. This material is partially based on work supported while serving at the National Science Foundation. Any opinion, findings, conclusions or recommendation expressed in this material are those of the author(s) and do not necessarily reflect the views of the National Science Foundation.

## APPENDIX

**Table 1A**. Summary of various machine learning based approaches used for the prediction of surface porosity in additive manufacturing.

| Machine Learning algorithm | AM Process | Material type | Defect type | Ground Truth | Dataset type | Features | Accuracy |
|---|---|---|---|---|---|---|---|
| Logistic regression and ANN [32] | DED | Ti-6Al-4V, H13 (Tool Steel) | Pores and cracks | Optical microscope and SEM images | Acoustic emission signals | Peak amplitude, kurtosis, rise time, duration, energy, peak amplitude frequency | MSE (LR) = 1.73 MSE (ANN) = 1.70 |
| Spectral CNN [33] | SLM | CL20ES stainless steel | Porosity | Optical Microscope images | Acoustic emission signals | Wavelet decomposition and energies of frequency bands | Acc = 89% |
| Process Monitoring [34] | SLM | Hastelloy X | Pores and microcracks | X-ray CT scans | Acoustic emission signals (Wireless transmission) | Signal characteristics in time / frequency domain | – |
| SVM [38] | SLM | SS 316L | Keyhole pores | X-ray radiography | Acoustic emission signals (FFT and EEMD) | Statistical features such as mean, kurtosis, skewness, etc., and energies of frequency bands | Acc > 93% |
| Tertiary-class CNN [39] | DED | Maraging Steel C300 | Cracks and Keyhole pores | Optical Microscope images | Acoustic emission signals | Signal characteristics in frequency and quefrency domain | Acc = 89% |
| Hybrid CNN [40] | DED | Maraging Steel C300 | Defects - Cracks and Keyholes | Optical Microscope images | Acoustic emission signals and Melt pool images | AE Signal spectrograms and melt pool images | Acc > 98% |
| K-Nearest Neighbors (clustering) [24] | DED | Ti-6Al-4V | Porosity | X-ray CT scans | Melt pool images | Melt pool geometry such as Circumference, Length, width, total area | Acc = 98.4% |
| Process Optimization Tool *DMP Meltpool* [25] | SLM | Ti-6Al-4V | Lack of fusion porosity | Density measurements per Archimedes principle and X-ray CT scans | Melt pool monitoring signals acquired by *DMP Melt pool* | – | Prediction sensitivity = 90% |
| Clustering via Self Organizing Maps (SOM) [26] | DED | Ti-6Al-4V | Porosity | X-ray CT scans to locate the pores | Melt pool images | Temperature distributions of the melt pool top surface and its geometry | Acc = 96% |

| Convolutional Neural Networks [27] | DED | Sponge Titanium powder | Porosity | X-ray CT scans | Melt pool images | Melt pool geometry such as total area, eccentricity and extent. | Acc = 91.2%. Volume prediction MSE = 1.32 |
|---|---|---|---|---|---|---|---|
| Residual Recurrent CNN [28] | DED | Ti-6Al-4V | Porosity | Pore size measured using X-ray CT scans and 3D reconstructions | HAZ images from pyrometer and infrared camera | Melt pool geometry and temperature distributions | Acc = 99.49% Inference time = 8.67 ms |
| Conv. Recurrent Neural Networks [29] | SLM | Ti-6Al-4V | Lack of fusion porosity | X-ray CT scans | Melt pool images of previous layer | Thermal Energy Planck (Temperature), Radiation (near-infrared region) | F1 score = 0.75 |
| 1D – CNN Classifier with Bayesian Optimization [30] | SLM | AISI 316L | Keyhole porosity | X-ray CT scans | Melt pool images | Melt pool geometry such as total area, eccentricity, perimeter and max / mean temperatures. | F1 score = 0.86 Recall = 0.87 Precision = 0.85 |
| Convolutional Neural Networks [11] | SLM | Ti-6Al-4V | Keyhole porosity | X-ray Imaging | Melt pool images | Wavelet decomposition (sectioned scalograms) | Acc = 100% |
| Long-term recurrent CNN [61] | DED | Ti-6Al-4V | Porosity | X-ray CT scans | Thermal Images from pyrometer & infrared camera | Pixels from melt pool images | Acc = 92.07% |
| Quadratic Discriminant Analysis (QDA) and SOM [5] | DED | Al-5083 | Porosity | Visual Inspection on Optical micrographs | Optical micrographs | Geometric features of pores such as total area of pore (pixels), eccentricity, solidity, perimeter, etc. | Acc = 95% |
| Random Forest [6] | DED | Al-5083 | Porosity | Visual inspection on optical micrographs: micropores, macropores and elongated pores | Optical micrographs | Geometry of pores from image processing such as perimeter, image texture, total area, circularity, etc. | Acc = 94.41% |
| Bayesian Classification [17] | SLM | Inconel 625 | Porosity | Visual Inspection of Camera Images | High resolution visual camera Images | Spectral features of image in the frequency domain. | Acc = 89.5% |
| Spatial Gaussian process model [9] | SLM | 17-4 PH stainless steel | Predict porosity | Density measured as per Archimedes principle (ASTM – B962-14) | SLM printing process parameters | Laser power and scan speed at various locations within a bounded region | Mean squared prediction error MSPE = 0.2593 |
| Multi-class CNN [36] | SLM | SS 316L | Keyholes, Lack of fusion and Conduction mode | Metallurgical Characterization | Laser power, scan speed, Normalized enthalpy) | Acoustic emission signal spectrograms | Classification Accuracy = 88% |
| Geometry – based simulation [23] | SLM | AlSi10Mg, Steel, TiAl6V4 | Lack of fusion porosity | Experimental density data | SLM printing process parameters | Melt pool shape, laser power, layer thickness, scan speed & hatch spacing | –<br>– |
| Bayesian Classifier [62] | SLM | SS316L,Co-Cr alloy, Ti-6Al-4V | Keyholes and Lack of fusion pores | X-ray CT scans | X-ray CT scans | Pore compactness, Aspect ratio & relative mean intensity - (Material and Process Agnostic Features) | Acc = 93% (same material) Acc = 90.2% (cross-material) |
| Prior – guided Neural Network approach [37] | SLM | Aluminum A205 | Localized porosity with SED < 50 µm | X-ray CT scans | – | InfiniAM Spectral Emission Sensor system data | Classification Accuracy = 73% |

**Key takeaways:**

1. Out of 23 studies, 13 primarily utilized thermal and optical images, and 5 employed Acoustic Emission (AE) signals for porosity prediction.
2. All except one study used a few hundred, and at most a thousand samples for both training and testing. Only one prior study [27] used a sufficiently large data (14,632 melt pool images) that can assure consistent prediction.
3. The prediction performance of all prior models are trained and tested under very restricted conditions. The thermal regimes that prevail during the printing of realistic geometries are not considered.
4. The highest spatial resolution reported in the earlier AE-based porosity prediction works was 4.8 mm. The present study can advance the state of the art by achieving a significantly finer spatial resolution of 0.5 mm.